\documentclass[aps,prd,epsfig,preprint,amsmath,amssymb,superscriptaddress,nofootinbib]{revtex4}
\usepackage{multirow}
\usepackage{graphicx,bm}
\usepackage[usenames]{color}
\usepackage{float,leftindex}
\usepackage{caption}
\usepackage{subcaption}
\usepackage{slashed}
\begin{document}
\draft
\title{Probing dimension-six SMEFT operators in heavy-quark sector at future muon collider}

\author{Serdar Spor}
\email[]{serdar.spor@beun.edu.tr}
\affiliation{Department of Medical Imaging Techniques, Zonguldak B\"{u}lent Ecevit University, 67100, Zonguldak, T\"{u}rkiye.}

\begin{abstract}

We investigate the sensitivity to the dimension-six Standard Model Effective Field Theory (SMEFT) operators which define the interactions of heavy-quarks with electroweak gauge bosons and with leptons via the process  $\mu^- \mu^+ \rightarrow Wtb$, followed by leptonic W-boson decays, at a 10 TeV muon collider with an integrated luminosity of 10 ab$^{-1}$. Signal and background events are generated at leading order using {\sc MadGraph5}$\_$aMC@NLO, interfaced with Pythia 8 for parton showering and hadronization, and passed through Delphes 3.5 for fast detector simulation.  The 95\% confidence level (C.L.) limits on nine Wilson coefficients are derived using a cut-based method for two $b$-tagging working points (70\% and 90\% efficiencies). These limits are significantly tighter than those currently reported by the CMS Collaboration; furthermore, these results highlight the advanced potential of a future high-energy muon collider to investigate new physics within the electroweak sector through precise measurements of heavy-quark interactions.

\end{abstract}


\maketitle

\section{Introduction} \label{Sec1}

The third-generation (top and bottom) quarks in the Standard Model (SM) of particle physics play a crucial role, due to their large masses, both in elucidating the nature of the electroweak symmetry breaking (EWSB) mechanism and in probing possible physical phenomena beyond the SM. The high precision measurements of the interactions of these heavy quarks, particularly with electroweak gauge bosons and leptons, provide a promising foundation for the search for new physics. In this regard, the Standard Model Effective Field Theory (SMEFT) offers a consistent methodological framework for parameterizing possible deviations from SM predictions in a model-independent and systematic manner. The SMEFT approach has been extensively studied in many areas of physics in recent years and has been widely applied to evaluate constraints on top-quark couplings \cite{Durieux:2018gwe,Durieux:2019gop,Bißmann:2021jhd,Blas:2022yec,Cornet-Gomez:2025edc,Machethe:2025yvc,Jiang:2025unz,Celada:2026rxz,Biswas:2026ged}. Multi-TeV muon colliders, planned to operate at energy levels beyond the top quark production threshold, are expected to significantly improve the current constraints on the electroweak interaction dynamics of the top quark. 

Muon colliders under design aim to accelerate and collide muon and antimuon beams at multi-TeV energy scales. In this study, we consider a muon collider configuration \cite{Accettura:2023kce} operating at a center-of-mass energy of 10 TeV with an integrated luminosity of 10 ab$^{-1}$. Muon collisions offer a remarkably clean experimental environment due to minimal pile-up, low hadronic noise, and well-defined initial-state kinematics, whereas proton-proton collisions at the Large Hadron Collider (LHC) take place in an environment dominated by QCD processes, high accumulation, and complex hadronic backgrounds. However, due to the short lifetime of muons, their decay products inevitably interact with beamline components, generating numerous secondary and tertiary particles that can reach the detector and give rise to beam-induced background (BIB), which poses a major experimental challenge for detector performance. To mitigate this effect, the BIB can be substantially suppressed by placing cone-shaped tungsten shielding nozzles along the beamline around the interaction point. Recent detector and simulation studies have shown encouraging results, indicating that the BIB can be successfully controlled \cite{Antonelli:2016ygs,Bartosik:2019edc,Bartosik:2020umx,Jindariani:2022wqw,Bartosik:2022ylz}. These capabilities at a muon collider enable precise reconstruction of final-state observables and significantly increase sensitivity to small deviations in two heavy quarks-gauge boson(s) and two heavy quarks-two leptons couplings via the process $\mu^-\mu^+ \rightarrow Wtb$. These deviations can arise from high-dimensional operators representing new physics effects that may emerge at high energy scales. Precise measurements of these couplings offer a powerful tool for searching for new physics signals in the electroweak sector of top and bottom quarks, complementing direct searches at multi-TeV colliders.

This study focuses specifically on dimension-six SMEFT operators that describe the couplings of two heavy quarks (top or bottom) - gauge boson(s) and two heavy quarks - two leptons. The primary motivation for searching for new physics in the top-quark sector is the exceptionally large mass of the top-quark and the resulting Yukawa couplings with the Higgs field, which are of approximately order one. Measurements to be performed at future high-energy muon colliders operating above the production threshold of the top quark pair are expected to significantly improve the current constraints on top-quark electroweak interactions and, in particular, on the operators that modify the interactions of two heavy quarks with gauge boson(s) and two heavy quarks with two leptons. In addition, the leptonic final state decay of the top quark arising from the process $\mu^-\mu^+ \rightarrow Wtb \rightarrow \ell^+ \nu_\ell b \ell^- \bar{\nu}_\ell \bar{b}$ provides an experimentally clean topology with relatively low background levels.

\section{The SMEFT in the Heavy Quark Sector} \label{Sec2}

The SMEFT provides a model-independent and consistent theoretical framework used to investigate the dynamics of new, heavy particles that lie beyond the direct kinematic reach of the Large Hadron Collider (LHC). It incorporates new physics effects through higher-dimensional effective operators while preserving the existing symmetry structures and field contents of the SM. The SMEFT Lagrangian, which extends the SM Lagrangian at the lowest order and suppresses high-dimensional operators relative to the energy scale ($\Lambda$), is expressed as follows:

\begin{eqnarray}
\label{eq.1} 
{\cal L}_{\text{eff}}={\cal L}_{\text{SM}}+\sum_{i}\frac{{C_i^{(d)}}}{\Lambda^{d-4}}{\cal O}_i^{(d)}
\end{eqnarray}

{\raggedright where ${\cal O}_i$ denotes the $i$th effective operator, ${C_i}^{(d)}$ is the $i$th Wilson coefficient and ${\cal L}_{\text{SM}}$ is the SM Lagrangian. The leading order contribution originates from the dimension-six $(d = 6)$ terms, which are taken into consideration in this study since odd-dimension operators violate lepton and baryon number conservation. The $\Lambda$ energy scale is fixed at 1 TeV in this study.}

In the search for new physics in top-quark production at muon collider, we investigate the interactions of two heavy quarks with gauge boson(s) and two heavy quarks with two leptons via the process $\mu^- \mu^+ \rightarrow Wtb$. These interactions are described by the dimension-six SMEFT operators listed in Table~\ref{tab1}, which are categorized into two categories: two-quark-gauge boson operators and two-quark-two-lepton operators, along with their corresponding Wilson coefficients \cite{Aguilar:2009unb,Grzadkowski:2010yws,Saavedra:2018ter}. In the notation used in this table, $i$, $j$, $k$, $l$ represent flavour indices; left-handed fermion doublets denoted by $q$ and $l$; right-handed fermion singlets denoted by $u$, $d$ and $e$. $\varphi$ is the Higgs doublet, $\varepsilon \equiv i \tau^2$ is the antisymmetric $SU(2)$ tensor and $\tau^I$ are the Pauli matrices.

\begin{table}[H]
\centering
\caption{SMEFT operators coupling two-heavy-quark to gauge boson(s) and two-heavy-quark to two-lepton, Wilson coefficient, their definition in terms of operator, and affected vertices. The energy scale is set to $\Lambda = 1$ TeV.}
\label{tab1}
\begin{tabular}{p{5.6cm}p{6.2cm}p{4.2cm}}
\hline \hline
\raggedright Two-quark-gauge boson operators & \raggedright Wilson coefficients $\equiv$ definition in terms of the operator & Affected vertices\\
\hline
$\leftindex^\ddagger {O}_{u\varphi}^{(ij)}=\bar{q}_i u_j \tilde{\varphi}(\varphi^\dagger \varphi)$ & $c_{t\varphi} \equiv C_{u\varphi}^{(33)}$ & $ttHHH$, $ttHH$, $ttH$.\\
$O_{\varphi q}^{1(ij)}=(\varphi^\dagger i\overleftrightarrow{D_\mu} \varphi)(\bar{q}_i \gamma^\mu q_j)$ & $c_{\varphi Q}^- \equiv C_{\varphi q}^{1(33)}-C_{\varphi q}^{3(33)}$ & $ttZHH$, $ttZH$, $ttZ$, $bbZHH$, $bbZH$, $bbZ$.\\
$O_{\varphi q}^{3(ij)}=(\varphi^\dagger i\overleftrightarrow{D^I_\mu} \varphi)(\bar{q}_i \gamma^\mu \tau^I q_j)$ & $c_{\varphi Q}^3\equiv C_{\varphi q}^{3(33)}$ & $tbWHH$, $tbWH$, $tbW$, $bbZHH$, $bbZH$, $bbZ$.\\
$O_{\varphi u}^{(ij)}=(\varphi^\dagger i\overleftrightarrow{D_\mu} \varphi)(\bar{u}_i \gamma^\mu u_j)$ & $c_{\varphi t}\equiv C_{\varphi u}^{(33)}$ & $ttZHH$, $ttZH$, $ttZ$.\\
$\leftindex^\ddagger {O}_{\varphi ud}^{(ij)}=(\tilde{\varphi}^\dagger i{D_\mu} \varphi)(\bar{u}_i \gamma^\mu d_j)$ & $c_{\varphi tb}\equiv C_{\varphi ud}^{(33)}$ & $tbWHH$, $tbWH$, $tbW$.\\
$\leftindex^\ddagger {O}_{uW}^{(ij)}=(\bar{q}_i \sigma^{\mu\nu} \tau^I u_j)(\tilde{\varphi}W_{\mu\nu}^I)$ & $c_{tW}\equiv C_{uW}^{(33)}$ & $tbWZH$, $tbW\gamma H$, $tbWH$, $tbW$, $ttWWH$, $ttWW$, $tt\gamma H$.\\
$\leftindex^\ddagger {O}_{dW}^{(ij)}=(\bar{q}_i \sigma^{\mu\nu} \tau^I d_j)(\varphi W_{\mu\nu}^I)$ & $c_{bW}\equiv C_{dW}^{(33)}$ & $tbWZH$, $tbW\gamma H$, $tbWH$, $tbWZ$, $tbW\gamma$, $tbW$, $bbWWH$, $bbWW$, $bbZH$, $bb\gamma H$, $bbZ$, $bb\gamma$.\\
$\leftindex^\ddagger {O}_{uB}^{(ij)}=(\bar{q}_i \sigma^{\mu\nu} u_j)(\tilde{\varphi} B_{\mu\nu})$ & $c_{tZ} \equiv -\text{sin}\theta_W C_{uB}^{(33)}+\text{cos}\theta_W C_{uW}^{(33)}$ & $ttZH$, $tt\gamma H$, $ttZ$, $tt\gamma$.\\ 
$\leftindex^\ddagger {O}_{uG}^{(ij)}=(\bar{q}_i \sigma^{\mu\nu} T^A u_j)(\tilde{\varphi} G^A_{\mu\nu})$ & $c_{tG} \equiv C_{uG}^{(33)}$ & $ttggH$, $ttgH$, $ttgg$,  $ttg$.\\
\hline
\raggedright Two-quark-two-lepton operators & &\\
\hline
$O_{lq}^{1(ijkl)}=(\bar{l}_i \gamma^{\mu} l_j)(\bar{q}_k \gamma^{\mu} q_l)$ & $c_{Ql}^{-(\ell)}\equiv C_{lq}^{1(\ell \ell33)}-C_{lq}^{3(\ell \ell33)}$ & $tt\nu_\mu \nu_\mu$, $bb\nu_\mu \nu_\mu$, $bb\mu \mu$.\\
$O_{lq}^{3(ijkl)}=(\bar{l}_i \gamma^{\mu}\tau^I l_j)(\bar{q}_k \gamma^{\mu}\tau^I q_l)$ & $c_{Ql}^{3(\ell)}\equiv C_{lq}^{3(\ell \ell33)}$ & $tt\nu_\mu \nu_\mu$, $tb\mu \nu_\mu$, $bb\mu \mu$.\\
$O_{lu}^{(ijkl)}=(\bar{l}_i \gamma^{\mu} l_j)(\bar{u}_k \gamma^{\mu} u_l)$ & $c_{tl}^{(\ell)}\equiv C_{lu}^{(\ell \ell33)}$ & $tt\nu_\mu \nu_\mu$.\\
$O_{eq}^{(ijkl)}=(\bar{e}_i \gamma^{\mu} e_j)(\bar{q}_k \gamma^{\mu} q_l)$ & $c_{Qe}^{(\ell)}\equiv C_{eq}^{(\ell \ell33)}$ & $tt\mu \mu$, $bb\mu \mu$.\\
$O_{eu}^{(ijkl)}=(\bar{e}_i \gamma^{\mu} e_j)(\bar{u}_k \gamma^{\mu} u_l)$ & $c_{te}^{(\ell)}\equiv C_{eu}^{(\ell \ell33)}$ & $tt\mu \mu$.\\
$\leftindex^\ddagger {O}_{lequ}^{1(ijkl)}=(\bar{l}_i e_j)\varepsilon(\bar{q}_k u_l)$ & $c_{t}^{S(\ell)}\equiv C_{lequ}^{1(\ell \ell33)}$ & $tb\mu \nu_\mu$, $tt\mu \mu$.\\
$\leftindex^\ddagger {O}_{lequ}^{3(ijkl)}=(\bar{l}_i \sigma^{\mu\nu} e_j)\varepsilon(\bar{q}_k \sigma_{\mu\nu} u_l)$ & $c_{t}^{T(\ell)}\equiv C_{lequ}^{3(\ell \ell33)}$ & $tb\mu \nu_\mu$, $tt\mu \mu$.\\
$\leftindex^\ddagger {O}_{ledq}^{(ijkl)}=(\bar{l}_i e_j)(\bar{d}_k q_l)$ & $c_{b}^{S(\ell)}\equiv C_{ledq}^{(\ell \ell33)}$ & $tb\mu \nu_\mu$, $bb\mu \mu$.\\
\hline \hline
\end{tabular}
\end{table}

{\raggedright where $B_{\mu\nu}$ and $W^I_{\mu\nu}$ are the field strength tensors of the electroweak interaction and a double dagger symbol is used to denote non-Hermitian operators. $D_\mu$ is the covariant derivative operator with $\varphi^\dagger i\overset\leftrightarrow{D}_\mu\varphi=\varphi^\dagger i(D_\mu \varphi)-i(D_\mu \varphi^\dagger)\varphi$, where $\varphi$ is the Higgs doublet in the SM with its dual $\tilde{\varphi}=i\tau^2 \varphi^*$.}

In this study, Monte Carlo simulations are employed to model both signal and background processes in order to investigate the effects of the dimension-six operators. We simulate the signal samples for the production ($\mu^-\mu^+ \rightarrow Wtb$) and decay ($\mu^-\mu^+ \rightarrow Wtb \rightarrow \ell^+ \nu_\ell b \ell^- \bar{\nu}_\ell \bar{b}$) modes and the relevant SM background processes at leading order using {\sc MadGraph5}$\_$aMC@NLO \cite{Alwall:2014cvc}.
The EFT framework is implemented via FeynRules \cite{Alloul:2014tfc} in the Universal FeynRules Output (UFO) \cite{Degrande:2012opq} format, specifically utilizing the dim6top UFO model \cite{Saavedra:2018ter}. Signal and background samples are interfaced with Pythia 8.3 \cite{Bierlich:2022uzx} for parton showering, fragmentation, and hadronization. Fast detector simulation is performed using Delphes 3.5.0 \cite{Favereau:2014wfb} configured with the default Muon Collider detector card.

\begin{figure}[H]
\centering
\includegraphics[scale=0.22]{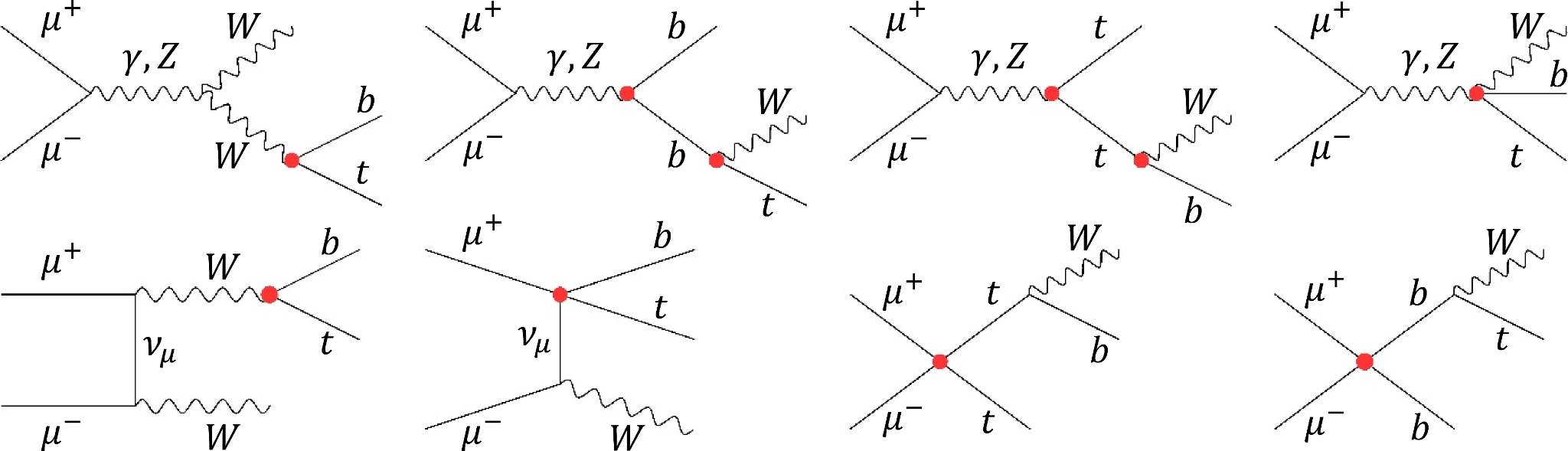}
\caption{The leading-order Feynman diagrams for the process $\mu^- \mu^+ \rightarrow Wtb$ involving anomalous vertices with two-heavy-quark-gauge boson(s) and two-heavy-quark-two-lepton.}
\label{fig1}
\end{figure}

The dim6top model \cite{Saavedra:2018ter}, built upon the Warsaw basis \cite{Grzadkowski:2010yws} of gauge-invariant dimension-six operators, focuses on operators involving third-generation quarks and enables tree-level simulations of their effects. This formalism includes a total of 26 operators classified into four main categories: four-heavy-quark operators, two-heavy-two-light-quark operators, two-heavy-quark-two-lepton operators, and two-heavy-quark-gauge boson(s) operators. Among these four categories, only the two-heavy-quark-two-lepton and two-heavy-quark-gauge boson(s) operators contribute at tree level to the signal process $\mu^- \mu^+ \rightarrow Wtb$ examined in this study. The leading-order Feynman diagrams containing the anomalous vertices that contribute to the signal process are illustrated in Fig.~\ref{fig1}. As shown in Fig.~\ref{fig1}, the anomalous vertices $tbW$, $bb\gamma$, $bbZ$, $tt\gamma$, $ttZ$, $tbW\gamma$, $tbWZ$, $tb\mu \nu_\mu$, $bb\mu \mu$, and $tt\mu \mu$ associated with the two-heavy-quark-two-lepton and two-heavy-quark-gauge boson(s) operators contribute to the signal process. The SMEFT operators and their corresponding Wilson coefficients for the two-heavy-quark-two-lepton and two-heavy-quark-gauge boson(s) interactions are presented in Table~\ref{tab1}. The SMEFT operators $\leftindex^\ddagger {O}_{u\varphi}^{(ij)}$, $\leftindex^\ddagger {O}_{uG}^{(ij)}$, $O_{eu}^{(ijkl)}$ and $O_{lu}^{(ijkl)}$ are not included in the anomalous vertices of the process $\mu^-\mu^+ \to Wtb$ in this study.

\begin{figure}[H]
\centering
\includegraphics[scale=0.2]{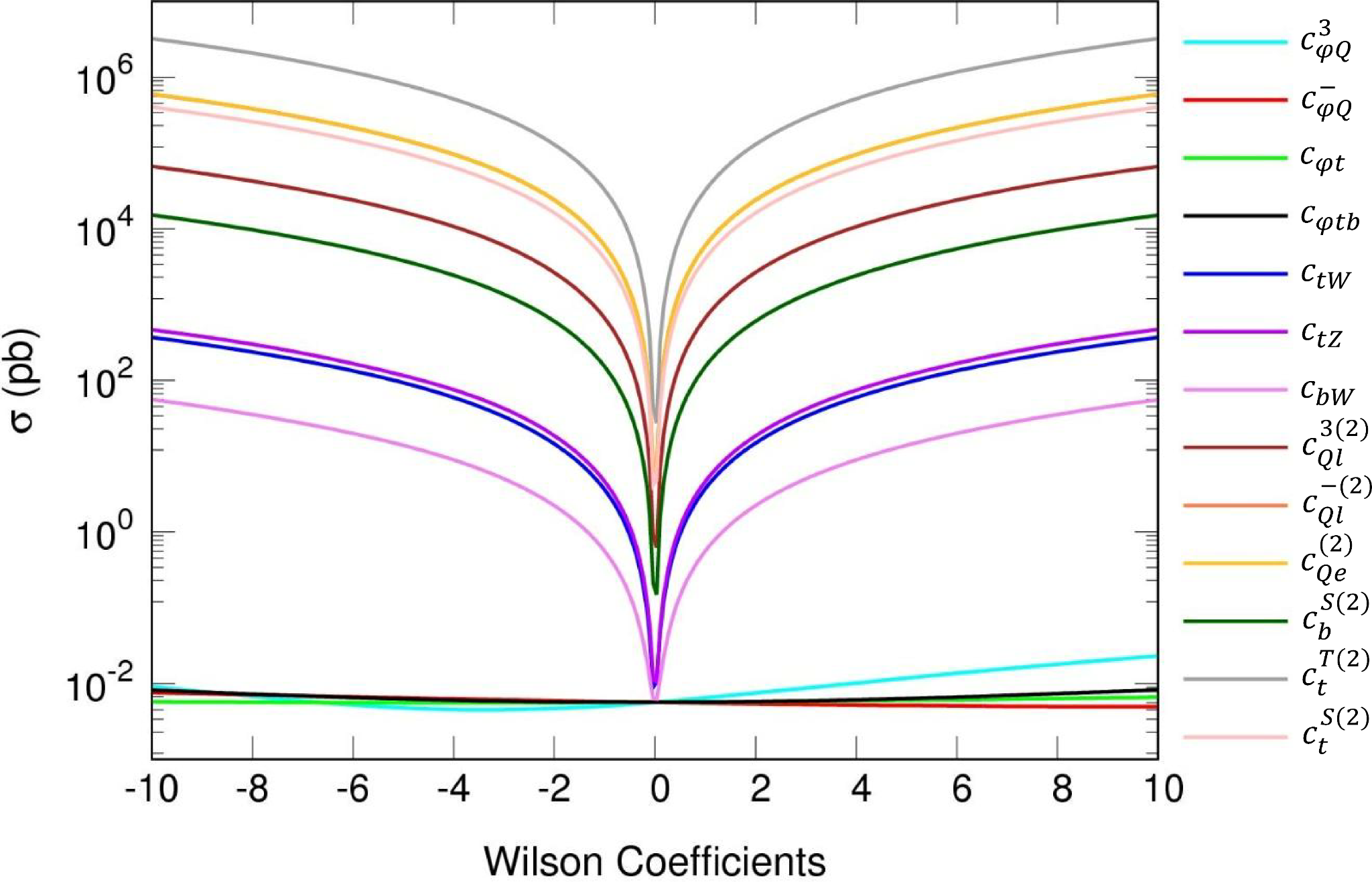}
\caption{The total cross-section of the process $\mu^- \mu^+ \rightarrow Wtb$ as a function of individual Wilson coefficients at a 10 TeV muon collider.}
\label{fig2}
\end{figure}

Fig.~\ref{fig2} displays the total cross-sections of the process $\mu^- \mu^+ \rightarrow Wtb$ as a function of each Wilson coefficient $c_{\varphi Q}^3$, $c_{\varphi Q}^-$, $c_{\varphi t}$, $c_{\varphi tb}$, $c_{tW}$, $c_{tZ}$, $c_{bW}$, $c_{Q \ell}^{3(2)}$, $c_{Q \ell}^{-(2)}$, $c_{Qe}^{(2)}$, $c_{b}^{S(2)}$, $c_{t}^{T(2)}$ and $c_{t}^{S(2)}$, varied individually with all others fixed to zero. Setting all Wilson coefficients to zero corresponds to the SM cross-section. The coefficients $c_{tW}$, $c_{tZ}$, $c_{bW}$, $c_{Q \ell}^{3(2)}$, $c_{Q \ell}^{-(2)}$, $c_{Qe}^{(2)}$, $c_{b}^{S(2)}$, $c_{t}^{T(2)}$ and $c_{t}^{S(2)}$ produce visibly larger cross-sections than those of the coefficients $c_{\varphi Q}^3$, $c_{\varphi Q}^-$, $c_{\varphi t}$ and $c_{\varphi tb}$. Therefore, we focus our subsequent sensitivity analyses on the nine coefficients: $c_{tW}$, $c_{tZ}$, $c_{bW}$, $c_{Q \ell}^{3(2)}$, $c_{Q \ell}^{-(2)}$, $c_{Qe}^{(2)}$, $c_{b}^{S(2)}$, $c_{t}^{T(2)}$ and $c_{t}^{S(2)}$.

\section{Analysis of Signal and Background Events} \label{Sec3}

In this analysis, processes that produce the same final state signatures as the signal process but are independent of the EFT operators are defined as the SM backgrounds. The final state of the signal process $\mu^- \mu^+ \rightarrow Wtb \rightarrow \ell\nu b\ell\nu b$ arises from the decay of the top-quark into a $W$ boson and a $b$-quark, followed by the leptonic decays of both $W$ bosons. To improve the signal sensitivity, a cut-based event selection strategy that exploits the characteristic signal kinematics is adopted.

Five SM background processes are taken into account: (i) The pure SM contribution to the same final state of the process $\mu^- \mu^+ \rightarrow Wtb \rightarrow \ell \nu b \ell \nu b$, where the top-quark decays into a $W$ boson and a $b$-quark, and both $W$ bosons decay leptonically (denoted as $Wtb$). (ii) The process $\mu^- \mu^+ \rightarrow HZZ \rightarrow bb\ell \ell \nu \nu$, where the Higgs boson decays into a $b\bar{b}$ pair and one $Z$-boson decays into a charged lepton pair, while the other $Z$-boson decays into a neutrino pair (denoted as $HZZ$). (iii) The process $\mu^- \mu^+ \rightarrow ZZZ \rightarrow bb\ell \ell \nu \nu$, where three $Z$-bosons decay into a $b\bar{b}$ pair, a charged lepton pair and a neutrino pair (denoted as $ZZZ$). (iv) The process $\mu^- \mu^+ \rightarrow HWW \rightarrow bb\ell \nu \ell \nu$, where the Higgs boson decays into a $b\bar{b}$ pair and both $W$-bosons decay leptonically (denoted as $HWW$). (v) The process $\mu^- \mu^+ \rightarrow ZWW \rightarrow bb\ell \nu \ell \nu$, where the $Z$-boson decays into a $b\bar{b}$ pair and the $W$-bosons decay leptonically (denoted as $ZWW$). For each signal and background sample, 500k events are generated in {\sc MadGraph5}$\_$aMC@NLO. Jets are reconstructed using the Valencia Linear Collider (VLC) algorithm \cite{Boronat:2015uev,Boronat:2018csf} operated in exclusive mode with a fixed number of jets $N_j = 2$, a cone size parameter $R = 1.0$ and parameters $\gamma=\beta=1$. Two $b$-tagging working points (WPs) defined in the Delphes Muon Collider card are investigated: the loose WP corresponding to a 90\% $b$-tagging efficiency and the medium WP corresponding to a 70\% $b$-tagging efficiency.

\begin{figure}[H]
\centering
\begin{subfigure}{0.47\linewidth}
\includegraphics[width=\linewidth]{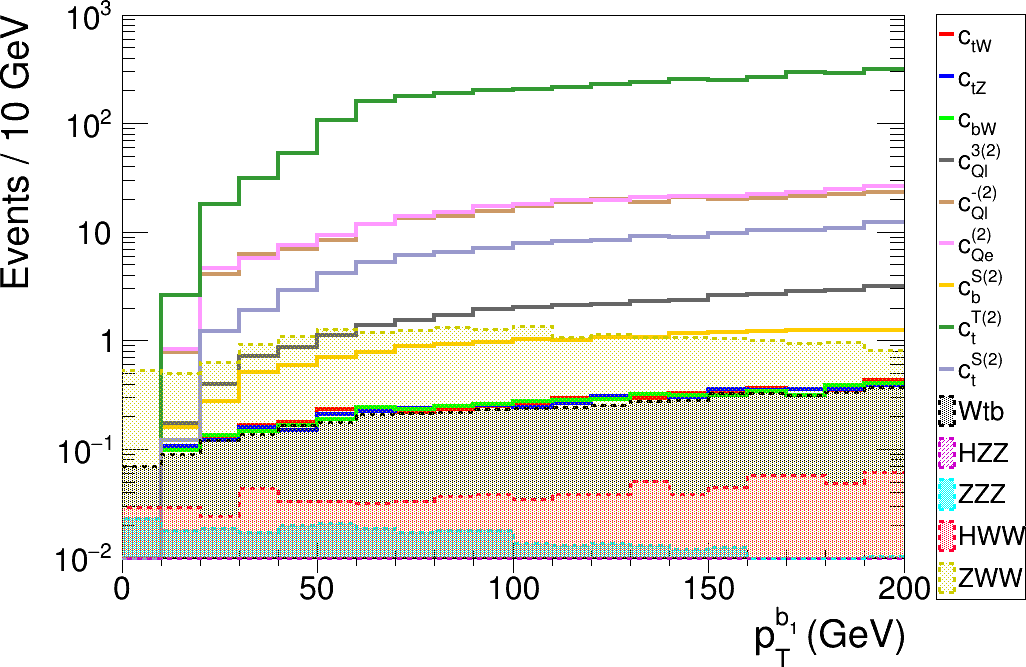}
\caption{}
\label{fig3:a}
\end{subfigure}\hfill
\begin{subfigure}{0.47\linewidth}
\includegraphics[width=\linewidth]{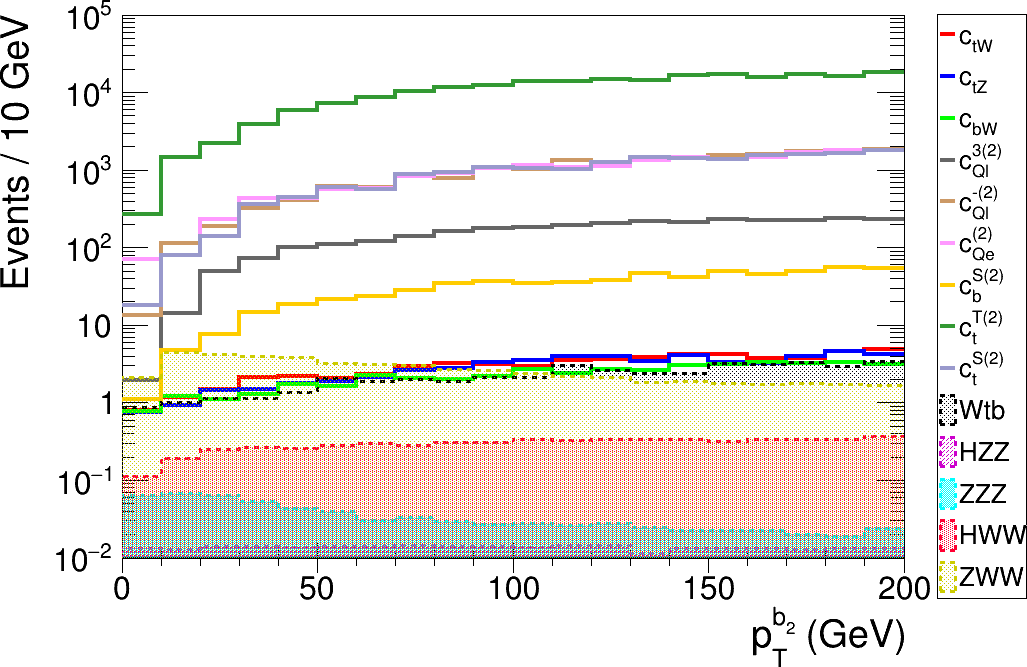}
\caption{}
\label{fig3:b}
\end{subfigure}\hfill

\begin{subfigure}{0.47\linewidth}
\includegraphics[width=\linewidth]{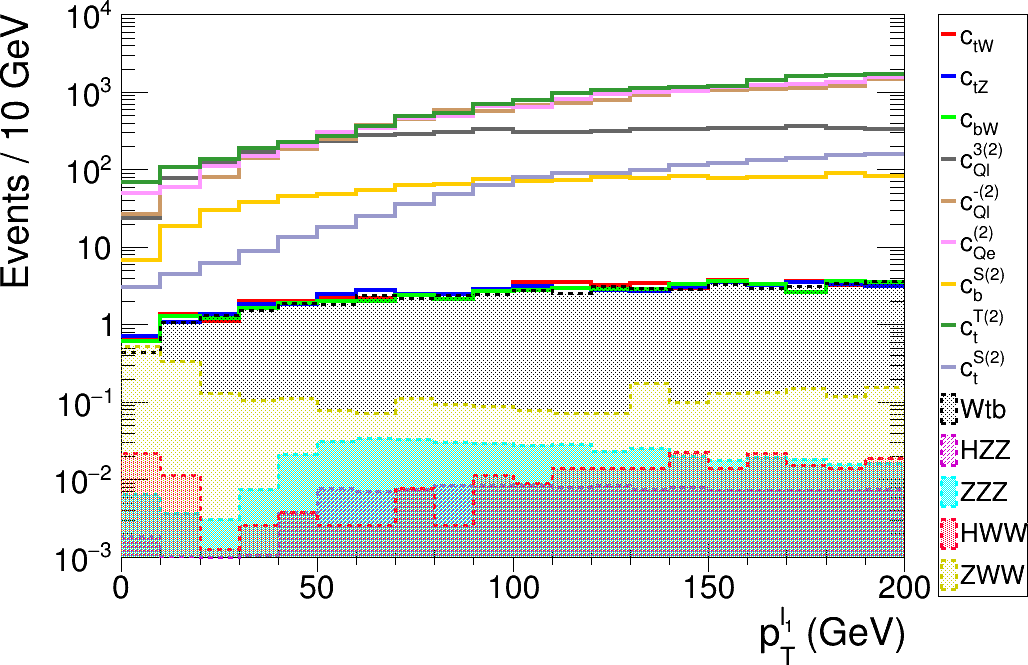}
\caption{}
\label{fig3:c}
\end{subfigure}\hfill
\begin{subfigure}{0.47\linewidth}
\includegraphics[width=\linewidth]{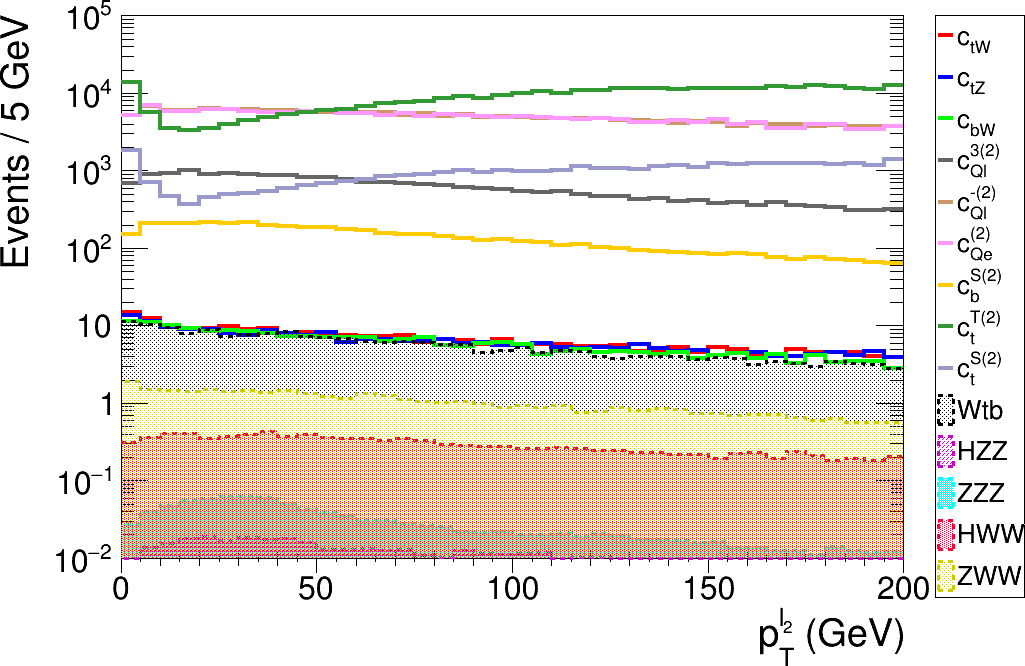}
\caption{}
\label{fig3:d}
\end{subfigure}\hfill

\caption{The transverse momentum distributions of the leading $b$-tagged jet (a), the sub-leading $b$-tagged jet (b), the leading charged lepton (c), and the sub-leading charged lepton (d) for the signal and background processes.}
\label{fig3}
\end{figure}

\begin{figure}[H]
\centering
\begin{subfigure}{0.48\linewidth}
\includegraphics[width=\linewidth]{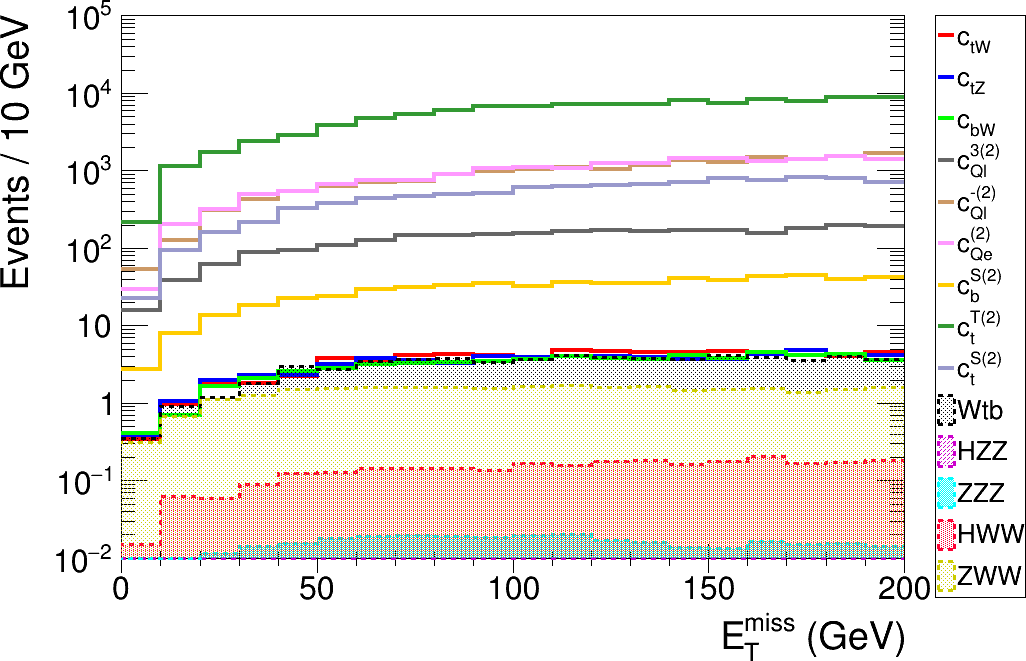}
\caption{}
\label{fig4:a}
\end{subfigure}\hfill
\begin{subfigure}{0.48\linewidth}
\includegraphics[width=\linewidth]{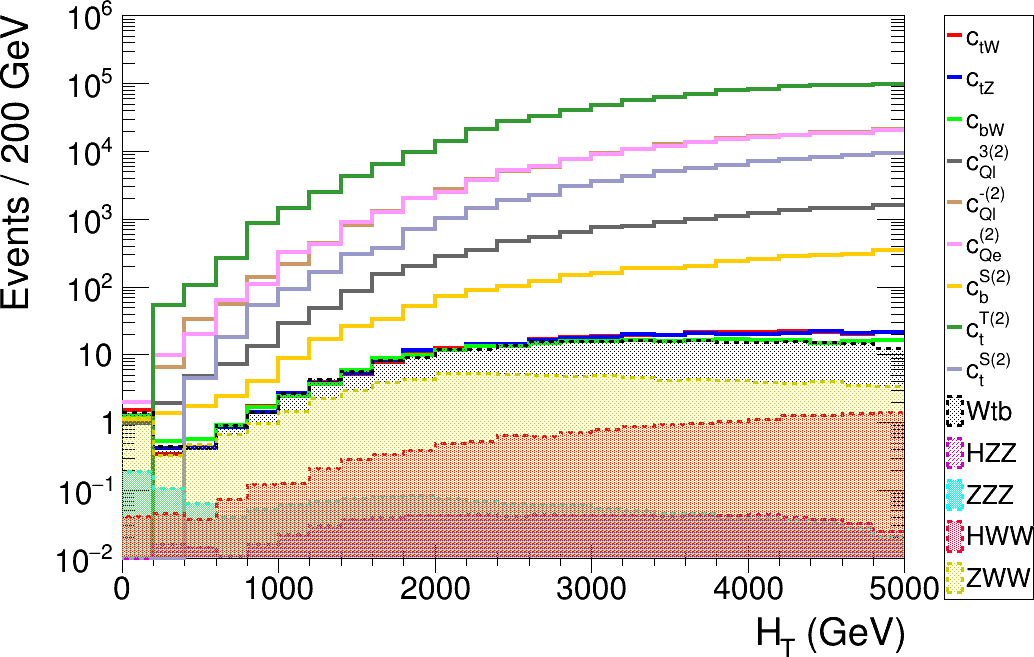}
\caption{}
\label{fig4:b}
\end{subfigure}\hfill

\caption{The distributions of the missing transverse energy (a) and the scalar sum of the transverse energy (b) for the signal and background processes.}
\label{fig4}
\end{figure}

\begin{figure}[H]
\centering
\begin{subfigure}{0.48\linewidth}
\includegraphics[width=\linewidth]{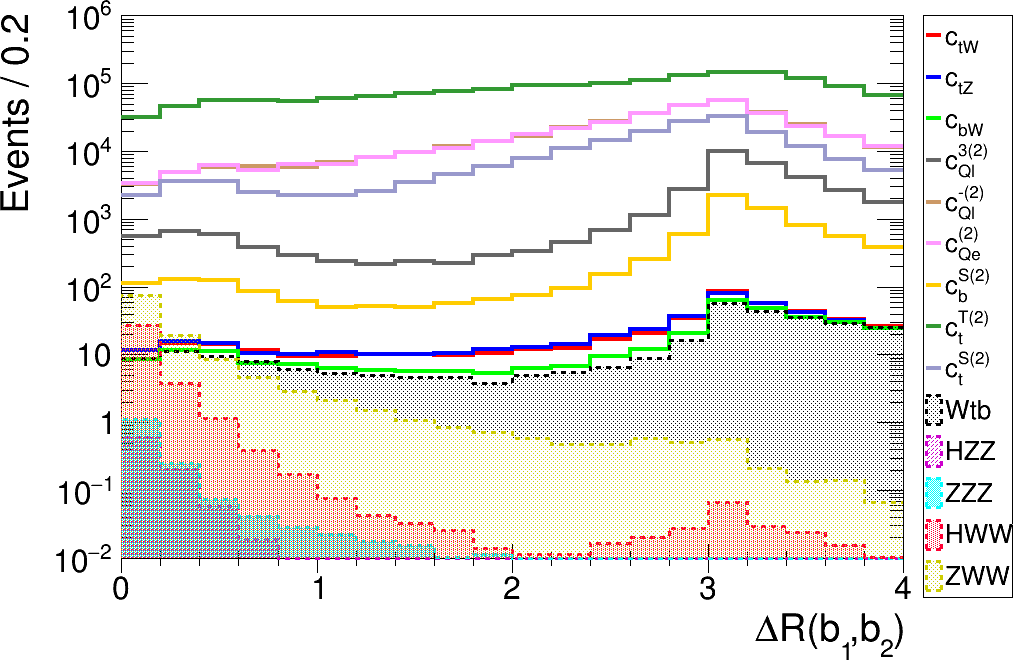}
\caption{}
\label{fig5:a}
\end{subfigure}\hfill
\begin{subfigure}{0.48\linewidth}
\includegraphics[width=\linewidth]{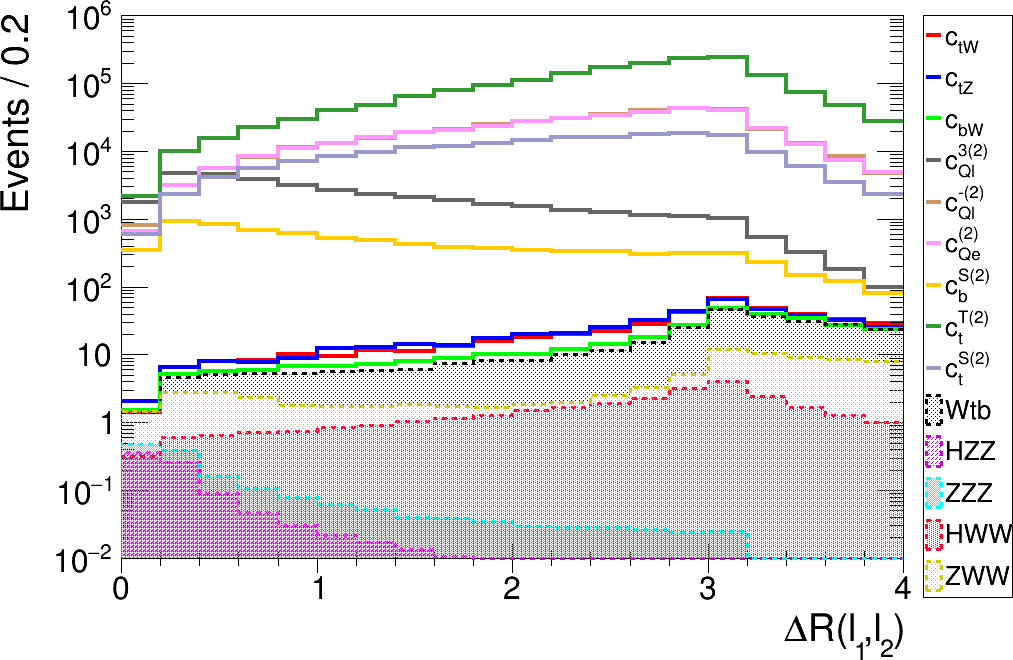}
\caption{}
\label{fig5:b}
\end{subfigure}\hfill
\begin{subfigure}{0.48\linewidth}
\includegraphics[width=\linewidth]{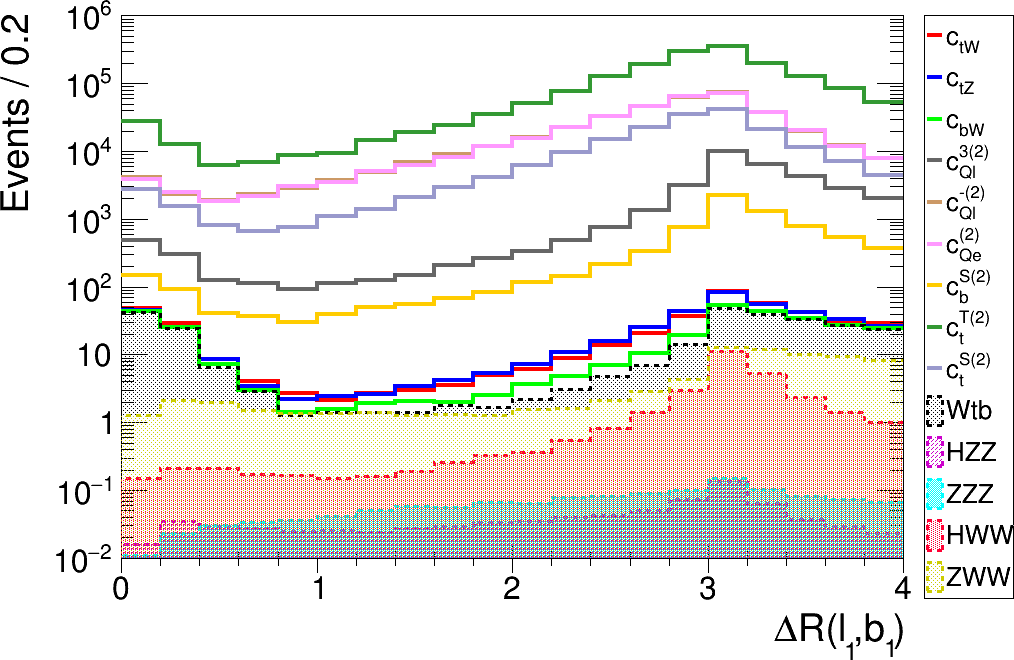}
\caption{}
\label{fig5:c}
\end{subfigure}\hfill

\caption{The angular distance distributions between the two b-tagged jets (a), between the two charged leptons (b), and between the leading charged lepton and the leading b-tagged jet (c) for the signal and background processes.}
\label{fig5}
\end{figure}

\begin{figure}[H]
\centering
\begin{subfigure}{0.48\linewidth}
\includegraphics[width=\linewidth]{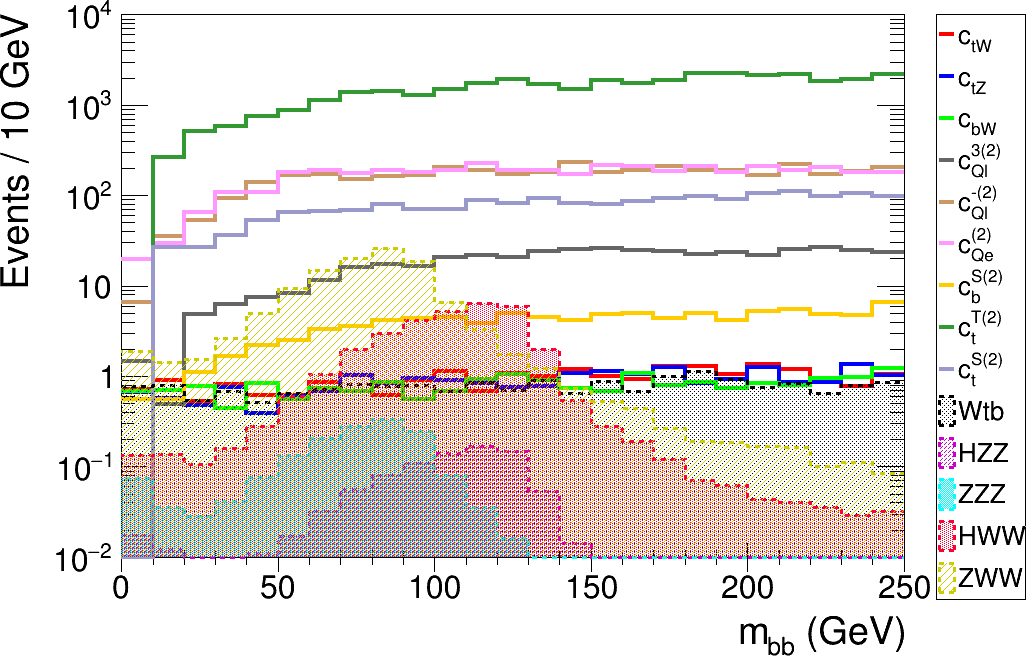}
\caption{}
\label{fig6:a}
\end{subfigure}\hfill
\begin{subfigure}{0.48\linewidth}
\includegraphics[width=\linewidth]{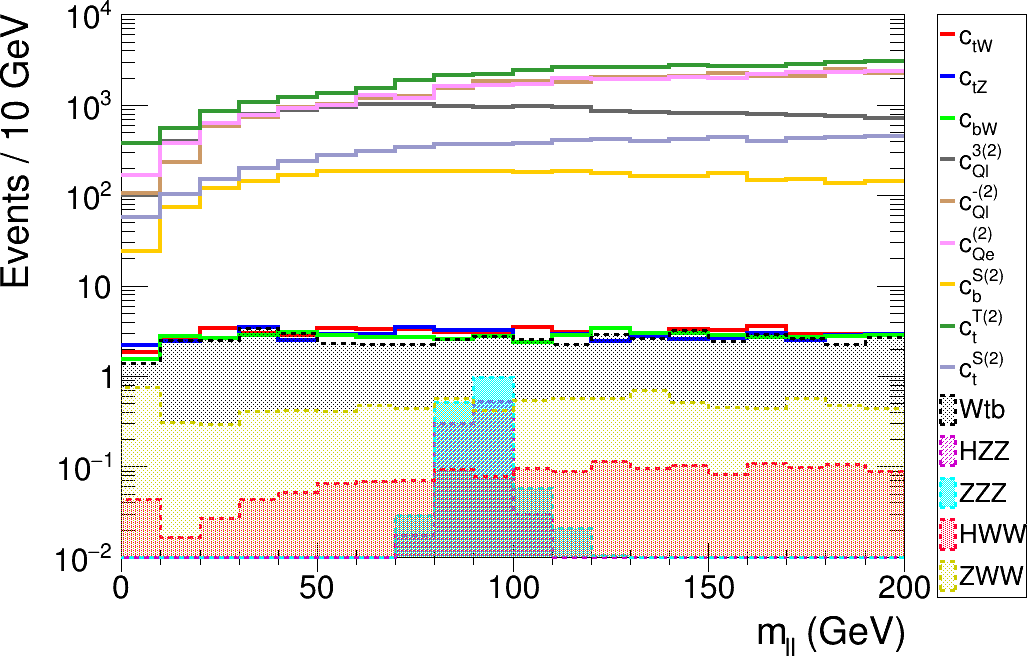}
\caption{}
\label{fig6:b}
\end{subfigure}\hfill

\begin{subfigure}{0.48\linewidth}
\includegraphics[width=\linewidth]{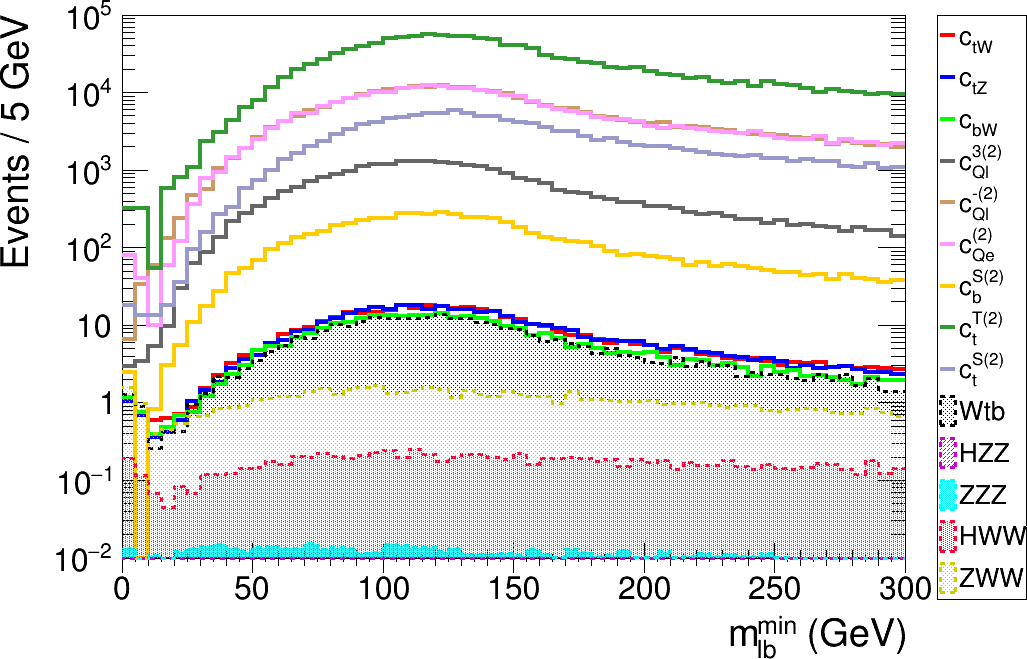}
\caption{}
\label{fig6:c}
\end{subfigure}\hfill

\caption{The invariant mass distributions of two b-tagged jets (a), charged lepton pair (b), charged lepton and b-tagged jet pair (c) for the signal and background processes.}
\label{fig6}
\end{figure}

The following kinematic cuts and requirements are applied to separate the signal from the background at a muon collider using cut-based analysis. As a pre-selection step (Cut-1), events are clustered into $N_{j}=2$ jets using the VLC algorithm with $R=1.0$, $\gamma=\beta=1$ and are required to contain $N_{b}=2$ b-tagged jets and $N_{\ell} \geq 2$ same-flavour opposite-sign charged leptons. The loose WP (90\% $b$-tagging efficiency) is used for all kinematic distributions in Figs.~\ref{fig3}-\ref{fig6}. Fig.~\ref{fig3} shows the transverse momentum distributions of the leading $b$-tagged jet $(b_1)$, the sub-leading $b$-tagged jet $(b_2)$, the leading charged lepton $(\ell_1)$, and the sub-leading charged lepton $(\ell_2)$ for the signal and the background processes. Based on these distributions, the signal can be distinguished from the background by $p_T^{b_1} > 20$ GeV, $p_T^{b_2} > 10$ GeV, $p_T^{\ell_1} > 40$ GeV, $p_T^{\ell_2} > 20$ GeV. Additionally, the pseudo-rapidities of the leading and sub-leading $b$-tagged jets ($\eta^{b_{1,2}}$), and the leading and sub-leading charged leptons ($\eta^{\ell_{1,2}}$) are required to satisfy $|\eta^{b_{1,2}}| < 2.5$ and $|\eta^{\ell_{1,2}}| < 2.5$, respectively, and these are referred to as Cut-2.

The distributions of the missing transverse energy ($E_T^{miss}$) and the scalar sum of the transverse energy ($H_T$) for the signal and background processes are shown in Fig.~\ref{fig4}. The angular distance of two particles in the $\eta – \phi$ space is defined as $\Delta R=\sqrt{|\Delta\eta|^2+|\Delta\phi|^2}$, where $\Delta\eta$ and $\Delta\phi$ are the differences in pseudo-rapidity and azimuthal angle, respectively. Fig.~\ref{fig5} illustrates the angular distance distributions between the two $b$-tagged jets, between the two charged leptons, and between the leading charged lepton and the leading $b$-tagged jet. Motivated by  Figs.~\ref{fig4}-\ref{fig5}, the kinematic cuts $E_T^{miss} > 30$ GeV, $H_T > 200$ GeV, $\Delta R_{bb} > 0.4$, $\Delta R_{\ell \ell} > 0.4$ and $\Delta R_{\ell b} > 0.4$ are imposed and referred to as Cut-3.

Fig.~\ref{fig6} presents the invariant mass distributions of the b-tagged jet pair, the charged lepton pair and the charged lepton - $b$-tagged jet pair. To suppress background originating from $H\rightarrow bb$ and $Z\rightarrow \ell \ell$ decays, we veto events using the invariant mass cuts on $|m_{bb}-m_H| > 10$ GeV and $|m_{\ell\ell}-m_Z| > 10$ GeV, where $m_H=125.1$ GeV and $m_Z=91.2$ GeV, along with the requirement $m_{\ell b} > 30$ GeV (Cut-4). All the kinematic cuts applied in the analysis are summarized in Table~\ref{tab2}.

\begin{table}[H]
\centering
\caption{Applied kinematic cuts used for the analysis at muon collider.}
\label{tab2}
\begin{tabular}{p{2cm}p{14cm}}
\hline \hline
Cut flow & Muon Collider at $\sqrt{s}=10$ TeV\\
\hline
Cut-1 & Jet clustering: Valencia (VLC) algorithm with $R=1.0$, \\
& $N_{b}=2$ (two b-tagged jets), $N_{\ell} \geq 2$ (at least two same flavor opposite-charge leptons)\\ \hline

Cut-2 & $p_T^{b_1} > 20$ GeV, $p_T^{b_2} > 10$ GeV, $p_T^{\ell_1} > 40$ GeV, $p_T^{\ell_2} > 20$ GeV,  \\
& $|\eta^{b_{1,2}}| < 2.5$, $|\eta^{\ell_{1,2}}| < 2.5$ \\ \hline
Cut-3 & $E_T^{miss} > 30$ GeV, \\
& $H_T > 200$ GeV, \\ 
& $\Delta R_{bb} > 0.4$, $\Delta R_{\ell \ell} > 0.4$, $\Delta R_{\ell b} > 0.4$ \\ \hline
Cut-4 & $|m_{bb}-m_H| > 10$ GeV, \\
& $|m_{\ell\ell}-m_Z| > 10$ GeV, \\  
& $m_{\ell b} > 30$ GeV \\

\hline \hline
\end{tabular}
\end{table}

The cumulative numbers of raw unweighted events ($N_{raw}$) for the signal ($c_{tW}$, $c_{tZ}$, $c_{bW}$, $c_{Q \ell}^{3(2)}$, $c_{Q \ell}^{-(2)}$, $c_{Qe}^{(2)}$, $c_{b}^{S(2)}$, $c_{t}^{T(2)}$ and $c_{t}^{S(2)}$) and the relevant backgrounds ($Wtb$, $HZZ$, $ZZZ$, $HWW$, $ZWW$) surviving after each kinematic cut are listed in Table~\ref{tab3} for both the 70\% and 90\% $b$-tagging efficiencies. The methodology involves assigning a value of $0.02$ to one of the coefficients while fixing the others at zero and each sample is generated with 500k events. In this table, the counts are raw simulated event numbers, not normalized to the integrated luminosity. 

The expected numbers of events $N_{exp}$ are calculated using the formula $N_{exp}=\sigma \times L_{int} \times \epsilon$. Cut efficiency ($\epsilon$) is precisely defined as the number of raw unweighted events surviving after a cut divided by the total number of events ($N_{raw}/N_{tot}$), where $N_{tot}=500k$. Depending on the cuts applied from Cut-1 to Cut-4, the signal-to-total-background ratios ($S/B_{tot}$) where $B_{tot}=B_{Wtb}+B_{HZZ}+B_{ZZZ}+B_{HWW}+B_{ZWW}$ increase for coefficients $c_{tW}$, $c_{tZ}$, $c_{bW}$, $c_{Q \ell}^{3(2)}$, $c_{Q \ell}^{-(2)}$, $c_{Qe}^{(2)}$, $c_{b}^{S(2)}$, $c_{t}^{T(2)}$ and $c_{t}^{S(2)}$ by factors of 1.38, 1.41, 1.30, 1.40, 1.98, 1.97, 1.44, 1.99 and 2.01, respectively, at 70\% b-tagging efficiency WP; and by factors of 1.34, 1.35, 1.28, 1.38, 1.92, 1.91, 1.39, 1.92 and 1.95, respectively, at 90\% b-tagging efficiency WP. This demonstrates that the applied cuts play a significant role in suppressing the relevant backgrounds.

\begin{table}[H]
\centering
\caption{The cumulative numbers of raw unweighted events for signal and relevant background processes surviving after kinematic cuts at muon collider.}
\label{tab3}
\begin{tabular}{p{1.2cm} p{1.2cm} p{1.4cm} p{1.4cm} p{1.4cm} p{1.4cm} p{1.4cm} p{1.4cm} p{1.4cm} p{1.4cm} p{1.2cm}}
\hline \hline
\multicolumn{11}{l}{Signals}  \\ \hline

Cuts & b-tag. & $c_{tW}$ & $c_{tZ}$ & $c_{bW}$ & $c_{Ql}^{3(2)}$ & $c_{Ql}^{-(2)}$ & $c_{Qe}^{(2)}$ & $c_b^{S(2)}$ & $c_t^{T(2)}$ & $c_t^{S(2)}$ \cr \hline

\multirow{2}{*}{Cut-1} & 70\% & 12930 & 13305 & 13690 & 15526 & 26443 & 25530 & 15135 & 18050 & 13593 \\ 
& 90\% & 22655 & 22865 & 23590 & 25926 & 44393 & 43685 & 25780 & 30065 & 22607  \\ \hline

\multirow{2}{*}{Cut-2} & 70\% & 10855 & 11420 & 11395 & 13108 & 24060 & 23105 & 12895 & 17425 & 13023 \\ 
& 90\% & 19020 & 19500 & 19615 & 22026 & 40480 & 39725 & 21750 & 29030 & 21673  \\ \hline

\multirow{2}{*}{Cut-3} & 70\% & 7525 & 7960 & 7570 & 9718 & 22093 & 21225 & 9675 & 15100 & 11487 \\ 
& 90\% & 13215 & 13515 & 13235 & 16462 & 37203 & 36355 & 16315 & 25085 & 19153  \\ \hline

\multirow{2}{*}{Cut-4} & 70\% & 7460 & 7890 & 7465 & 9082 & 21907 & 21060 & 9160 & 15035 & 11447 \\ 
& 90\% & 13075 & 13375 & 13070 & 15498 & 36840 & 36105 & 15485 & 24935 & 19073 \\ 
\hline
\multicolumn{7}{l}{Backgrounds} & & & &  \\ \hline

Cuts & b-tag. & $Wtb$ & $HZZ$ & $ZZZ$ & $HWW$ & $ZWW$ & & \cr \hline

\multirow{2}{*}{Cut-1} & 70\% & 12840 & 47760 & 45890 & 19560 & 12455 & & \\ 
& 90\% & 22592 & 79319 & 76700 & 32155 & 20560 & &  \\ \hline

\multirow{2}{*}{Cut-2} & 70\% & 10640 & 43810 & 38465 & 17810 & 10440 &  &\\ 
& 90\% & 18855 & 72530 & 63520 & 29010 & 17095 &  &  \\ \hline

\multirow{2}{*}{Cut-3} & 70\% & 6795 & 544 & 1305 & 925 & 1870 &  &\\ 
& 90\% & 12195 & 944 & 2210 & 1535 & 3340 &  &  \\ \hline

\multirow{2}{*}{Cut-4} & 70\% & 6720 & 35 & 105 & 685 & 1780 &  &  \\ 
& 90\% & 12010 & 59 & 170 & 1085 & 3130 &  &  \\  

\hline \hline
\end{tabular}
\end{table}

\section{Sensitivities on the Wilson Coefficients} \label{Sec4}

The $\chi^2$ test is used to calculate the sensitivity of the top quark’s interactions with leptons, bosons, and other heavy quarks (top or bottom) in the process $\mu^- \mu^+ \rightarrow Wtb$. The analysis utilizes the $\chi^2$ distribution to determine limits at 95\% confidence level (C.L.) with a threshold value of $\chi^2=3.84$ corresponding to one degree of freedom. The $\chi^2$ function is defined as follows:

\begin{eqnarray}
\label{eq.12}
\chi^{2}=\sum_{i}^{n_{bins}} (\frac{N_{i}^{NP}-N_{i}^{B}}{N_{i}^{B}\Delta_{i}})^{2}
\end{eqnarray}

\begin{figure}[H]
\centering
\begin{subfigure}{0.48\linewidth}
\includegraphics[width=\linewidth]{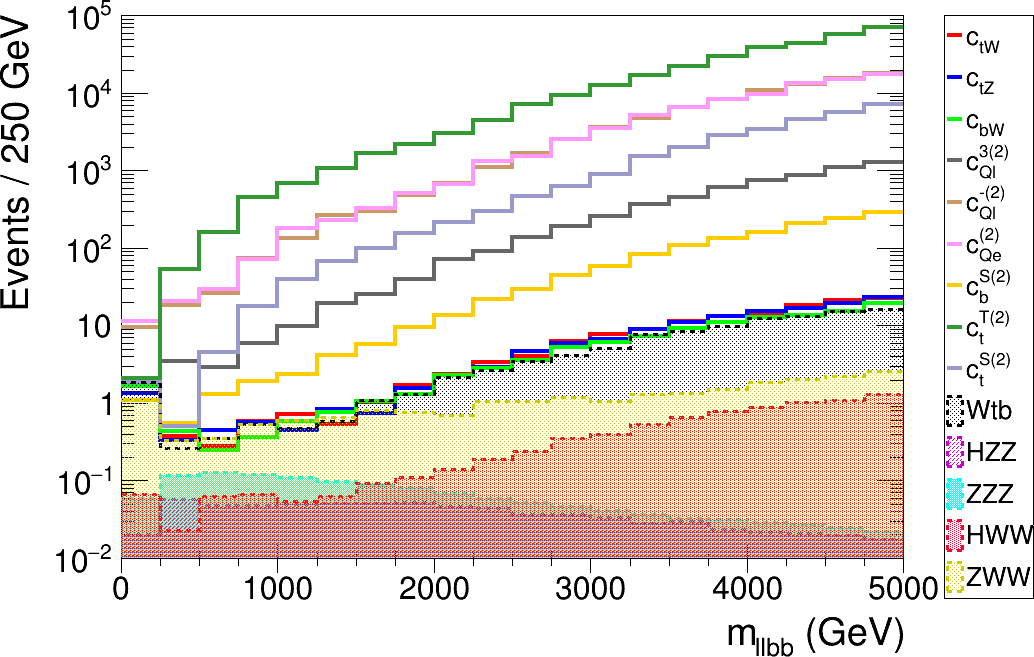}
\caption{}
\label{fig7:a}
\end{subfigure}\hfill
\begin{subfigure}{0.48\linewidth}
\includegraphics[width=\linewidth]{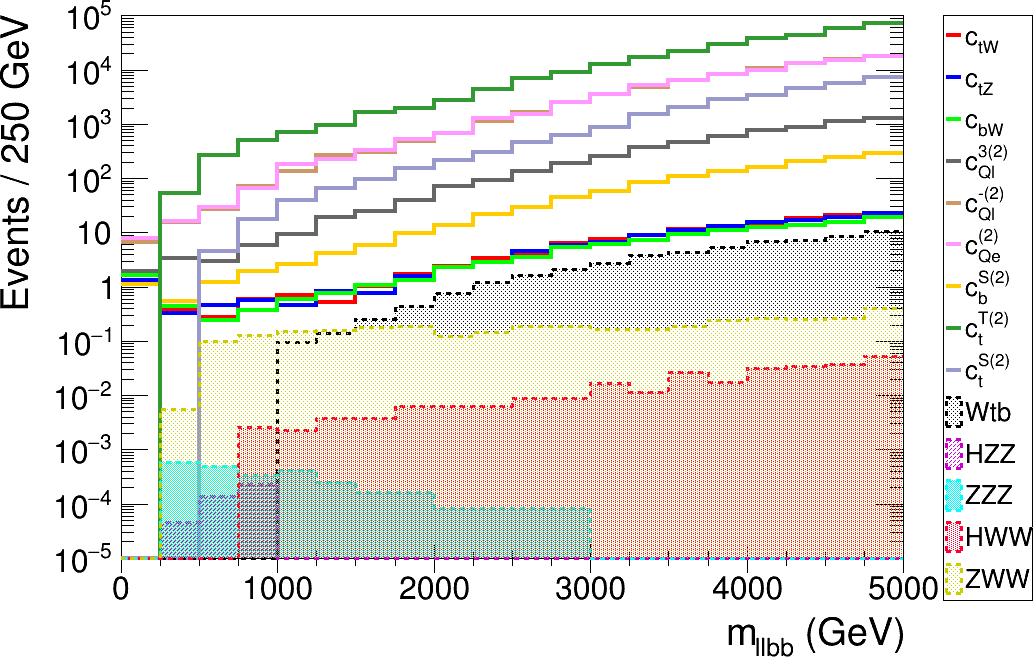}
\caption{}
\label{fig7:b}
\end{subfigure}\hfill

\caption{The invariant mass distributions of the two-lepton and two-$b$-jet system after Cut-1 (a) and after Cut-4 (b) for the signal and relevant background processes.}
\label{fig7}
\end{figure}

{\raggedright where $N_{i}^{B}$ is the number of events of relevant total backgrounds in ith bin and $N_{i}^{NP}$ is the total number of events including contributions of effective couplings and relevant total backgrounds in ith bin. These events are calculated by integrating the invariant mass $m_{\ell \ell bb}$ distribution of $\ell\ell bb$ system after Cut-4. Here, $\Delta_{i}=\sqrt{1/N_i^B}$ represents the statistical uncertainty in each bin, with events normalized to an integrated luminosity of 10 ab$^{-1}$ at $\sqrt{s}=10$ TeV. Fig.~\ref{fig7} shows the four-body invariant mass distributions of the two-lepton and two-$b$-jet system after Cut-1 (left) and after Cut-4 (right) for the signal and relevant background processes at muon collider.}

\begin{figure}[H]
\centering
\includegraphics[scale=0.4]{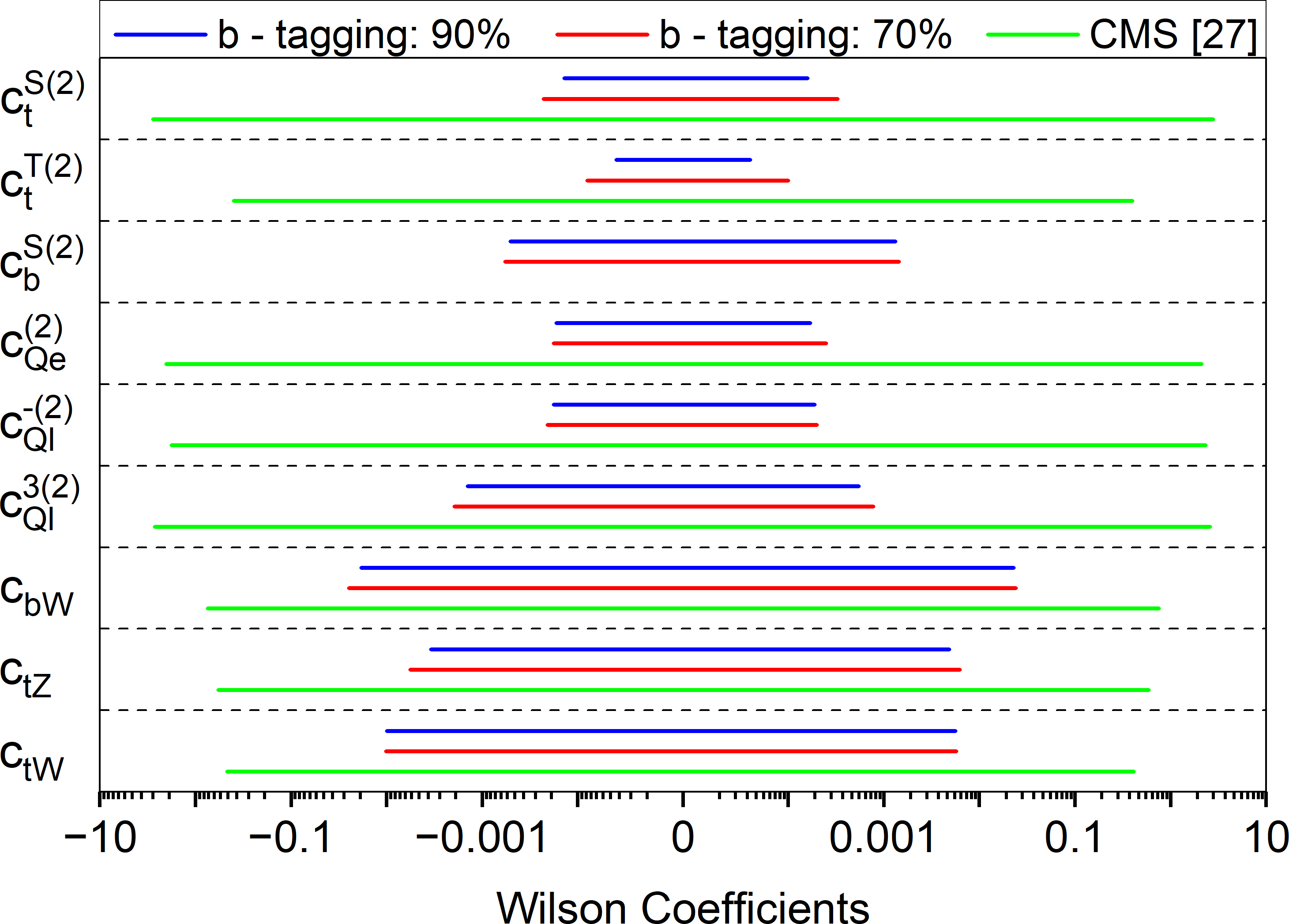}
\caption{Summary of 95\% C.L. intervals for the Wilson coefficients $c_{tW}$, $c_{tZ}$,$c_{bW}$, $c_{Ql}^{3(2)}$, $c_{Ql}^{-(2)}$, $c_{Qe}^{(2)}$, $c_b^{S(2)}$, $c_t^{T(2)}$ and $c_t^{S(2)}$ for 70\% and 90\% b-tagging efficiencies.}
\label{fig8}
\end{figure}

Table~\ref{tab4} summarizes the 95\% C.L. limits on the coefficients  $c_{tW}$, $c_{tZ}$, $c_{bW}$, $c_{Q \ell}^{3(2)}$, $c_{Q \ell}^{-(2)}$, $c_{Qe}^{(2)}$, $c_{b}^{S(2)}$, $c_{t}^{T(2)}$ and $c_{t}^{S(2)}$ obtained from a single-parameter analysis for both the 70\% and 90\% $b$-tagging working points at a 10 TeV muon collider with ${\cal L}_{\text{int}}=10$ ab$^{-1}$, alongside the limits from present experimental studies. As illustrated in Fig.~\ref{fig8}, the sensitivity of the limits for the muon collider improves noticeably as the $b$-tagging efficiency increases, and a better limit is achieved at the loose working point (90\% b-tagging efficiency). Furthermore, comparing the limits of recent experimental studies with our own limits, as shown in Table~\ref{tab4}, is one of the notable aims of our study, and a detailed comparison of these limits will be presented in Section~\ref{Sec5}.

\begin{table}[H]
\centering
\caption{The 95\% C.L. limits on the Wilson coefficients obtained in our study and in various experimental studies.}
\label{tab4}
\begin{tabular}{p{1.8cm}p{0.1cm}p{3.4cm}p{3.2cm}p{0.1cm}p{4.7cm}p{2cm}}
\hline \hline
&& \multicolumn{2}{c}{This study} && \multicolumn{2}{c}{Experimental studies (other WCs fixed to SM)} \\
\cline{3-4} \cline{6-7} 
Coefficients &&  \multicolumn{1}{c}{$b$-$tag.=70\%$} &  \multicolumn{1}{c}{$b$-$tag.=90\%$} &&  \multicolumn{1}{l}{Ref.~\cite{Sirunyan:2021rfv}} &  \multicolumn{1}{l}{Ref.~\cite{Hayrapetyan:2023yjz}} \\
\hline 
$c_{tW}$ &&  $[-0.01024; 0.00573]$ & $[-0.01005; 0.00560]$ && $[-2.15; -0.29]\cup[0.21; 1.96]$ & $[-0.47; 0.41]$ \\
$c_{tZ}$ && $[-0.00572; 0.00627]$ & $[-0.00347; 0.00486]$ && $[-2.14; 2.19]$ & $[-0.58; 0.59]$ \\
$c_{bW}$ && $[-0.02505; 0.02418]$ & $[-0.01848; 0.02271]$ && $[-4.12; 4.09]$ & $[-0.75; 0.75]$ \\
$c_{Ql}^{3(2)}$ && $[-0.00196; 0.00078]$ & $[-0.00143; 0.00055]$ && $[-9.91; 9.5]$ & $[-2.69; 2.58]$ \\ 
$c_{Ql}^{-(2)}$ && $[-0.00021; 0.00020]$ & $[-0.00018; 0.00019]$ && $[-4.76; 5.83]$ & $[-1.8; 2.33]$ \\
$c_{Qe}^{(2)}$ && $[-0.00018; 0.00025]$ & $[-0.00017; 0.00017]$ && $[-5.2; 5.36]$ & $[-2.04; 2.12]$ \\
$c_b^{S(2)}$ && $[-0.00058; 0.00144]$ & $[-0.00051; 0.00132]$ && - & - \\
$c_t^{T(2)}$ && $[-0.00008; 0.00010]$ & $[-0.00004; 0.00004]$ && $[-1.01; 1.01]$ & $[-0.4; 0.4]$ \\
$c_t^{S(2)}$ && $[-0.00023; 0.00033]$ & $[-0.00014; 0.00016]$ && $[-7.7; 7.7]$ & $[-2.8; 2.8]$ \\
\hline \hline
\end{tabular}
\end{table}

To provide an approximate assessment of the validity of the dimension-six EFT framework, under the condition that the characteristic energy scale of the hard scattering process remains below the scale of new physics ($E < \Lambda_{\text{NP}}$), we employ power-counting relations \cite{Contino:2016tlz,Dhondt:2018ybq}. It is assumed that the Wilson coefficients for four-fermion operators scale as $C_i/\Lambda^2 \sim g_*^2/\Lambda_{\text{NP}}^2$, while those for electroweak dipole operators as $C_i/\Lambda^2 \sim g_V g_*/\Lambda_{\text{NP}}^2$. Here, $g_*$ represents the characteristic coupling constant of the new physics sector, and $g_V$ denotes the relevant electroweak gauge coupling constant ($g_V = g_W \approx 0.653$ and $g_V = g_Z \approx 0.743$). Taking $g_* = 4\pi$ as a reference upper bound for the new physics coupling, these power-counting relations yield estimates for the maximum new physics scale: $\Lambda_{\text{NP},i}^{\max} \sim (4\pi/\sqrt{|C_i|})\,\Lambda$ for four-fermion operators, and $\Lambda_{\text{NP},i}^{\max} \sim \sqrt{4\pi g_V/|C_i|}\,\Lambda$ for dipole operators. Using the most precise 95\% C.L. limits of the coefficients obtained at the 90\% b-tagging working point, and with $\Lambda = 1$ TeV taken as the reference normalization, the estimated upper scales for four-fermion operators are found to be approximately 332.3 TeV for $c_{Ql}^{3(2)}$, 911.7 TeV for $c_{Ql}^{-(2)}$, 963.8 TeV for $c_{Qe}^{(2)}$, 345.9 TeV for $c_b^{S(2)}$, 1986.9 TeV for $c_t^{T(2)}$, and 993.5 TeV for $c_t^{S(2)}$. For the dipole operators, under the assumption of electroweak coupling normalization, the estimated upper scales are approximately 28.6 TeV for $c_{tW}$, 43.8 TeV for $c_{tZ}$, and 19.0 TeV for $c_{bW}$. Although these maximum scale estimates exceed the 10 TeV collider energy, this comparison should be interpreted as a conditional consistency check based on power counting, rather than a definitive demonstration of the validity of the EFT. In particular, smaller values of $g_*$, additional loop or chirality suppressions, and contributions from higher-dimensional operators could reduce the range of validity for the EFT. Consequently, these estimates suggest that the obtained limits on the Wilson coefficients could be consistent with an EFT interpretation at a 10 TeV muon collider, under the specified assumptions of a strongly coupled UV completion.

\begin{figure}[H]
\centering
\begin{subfigure}{0.32\linewidth}
\includegraphics[width=\linewidth]{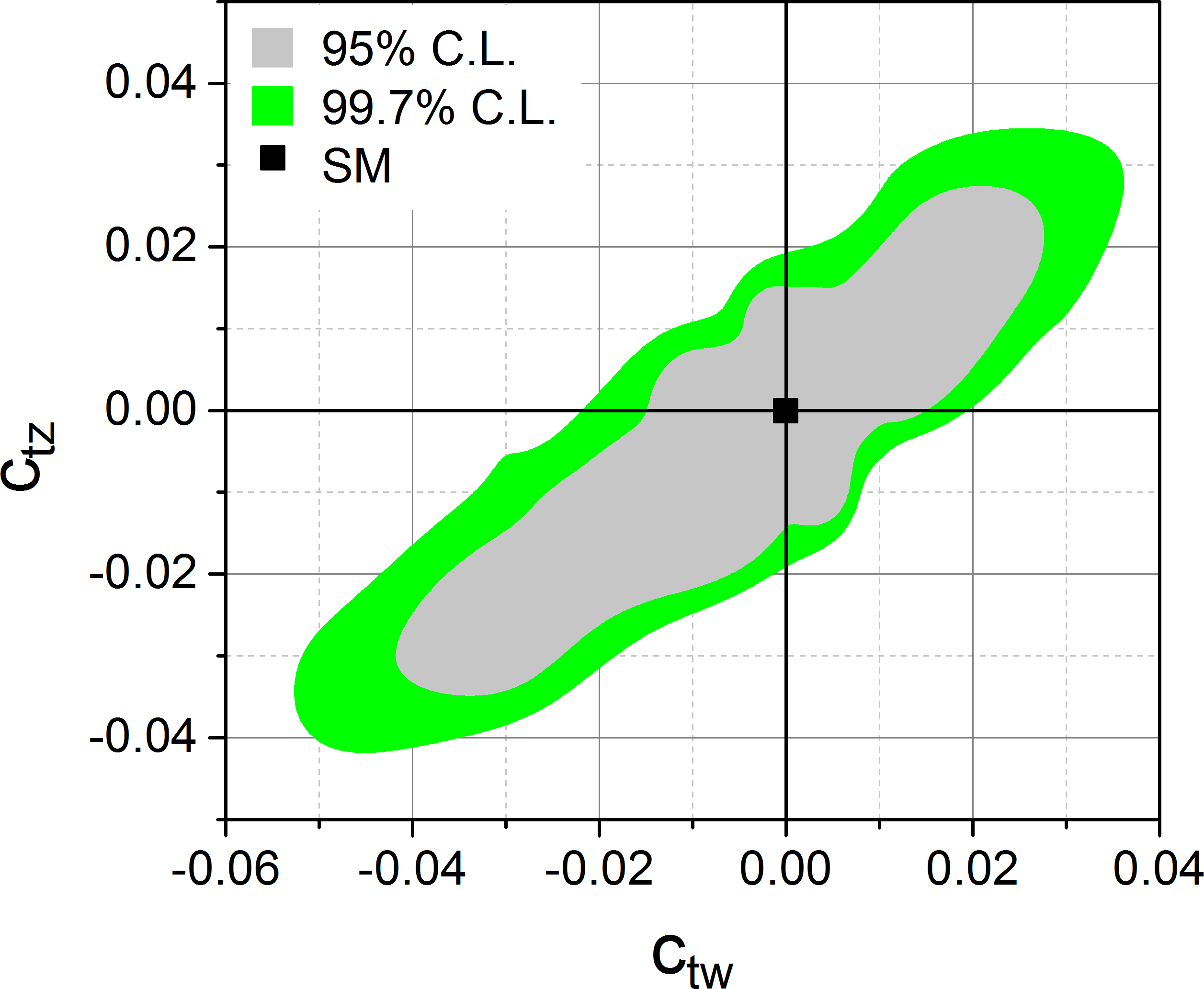}
\caption{}
\label{fig9:a}
\end{subfigure}\hfill
\begin{subfigure}{0.31\linewidth}
\includegraphics[width=\linewidth]{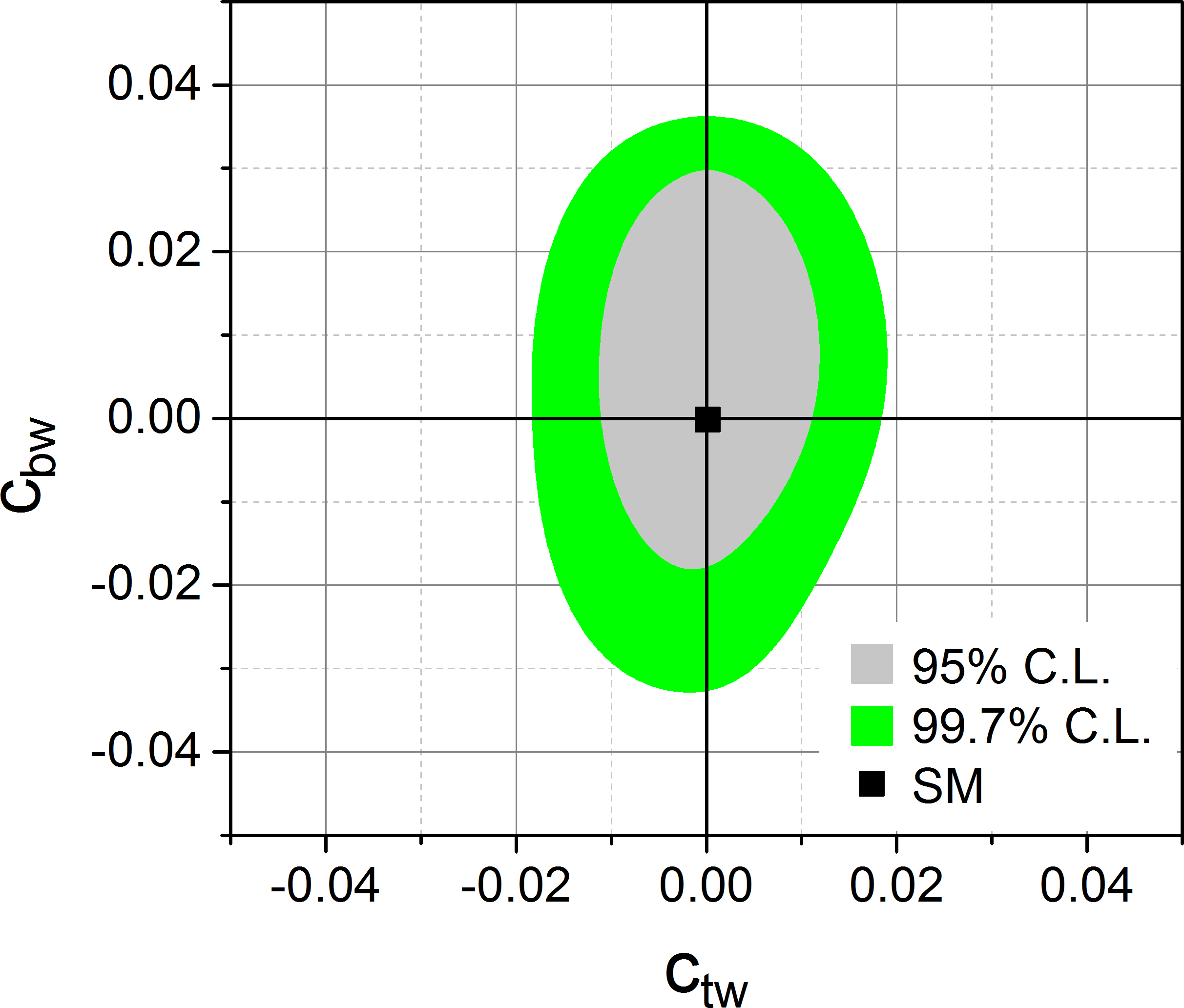}
\caption{}
\label{fig9:b}
\end{subfigure}\hfill
\begin{subfigure}{0.34\linewidth}
\includegraphics[width=\linewidth]{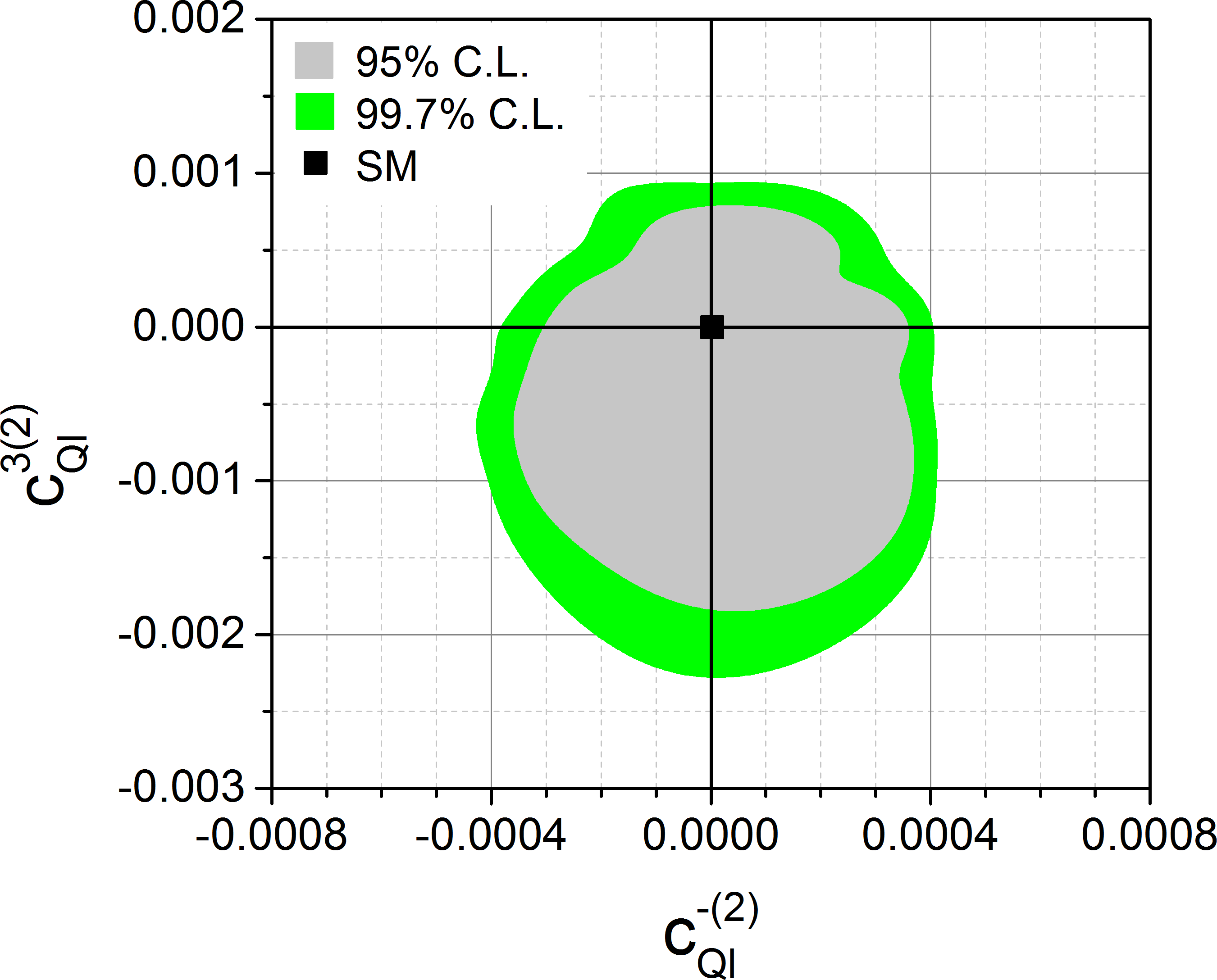}
\caption{}
\label{fig9:c}
\end{subfigure}\hfill

\centering
\begin{subfigure}{0.33\linewidth}
\includegraphics[width=\linewidth]{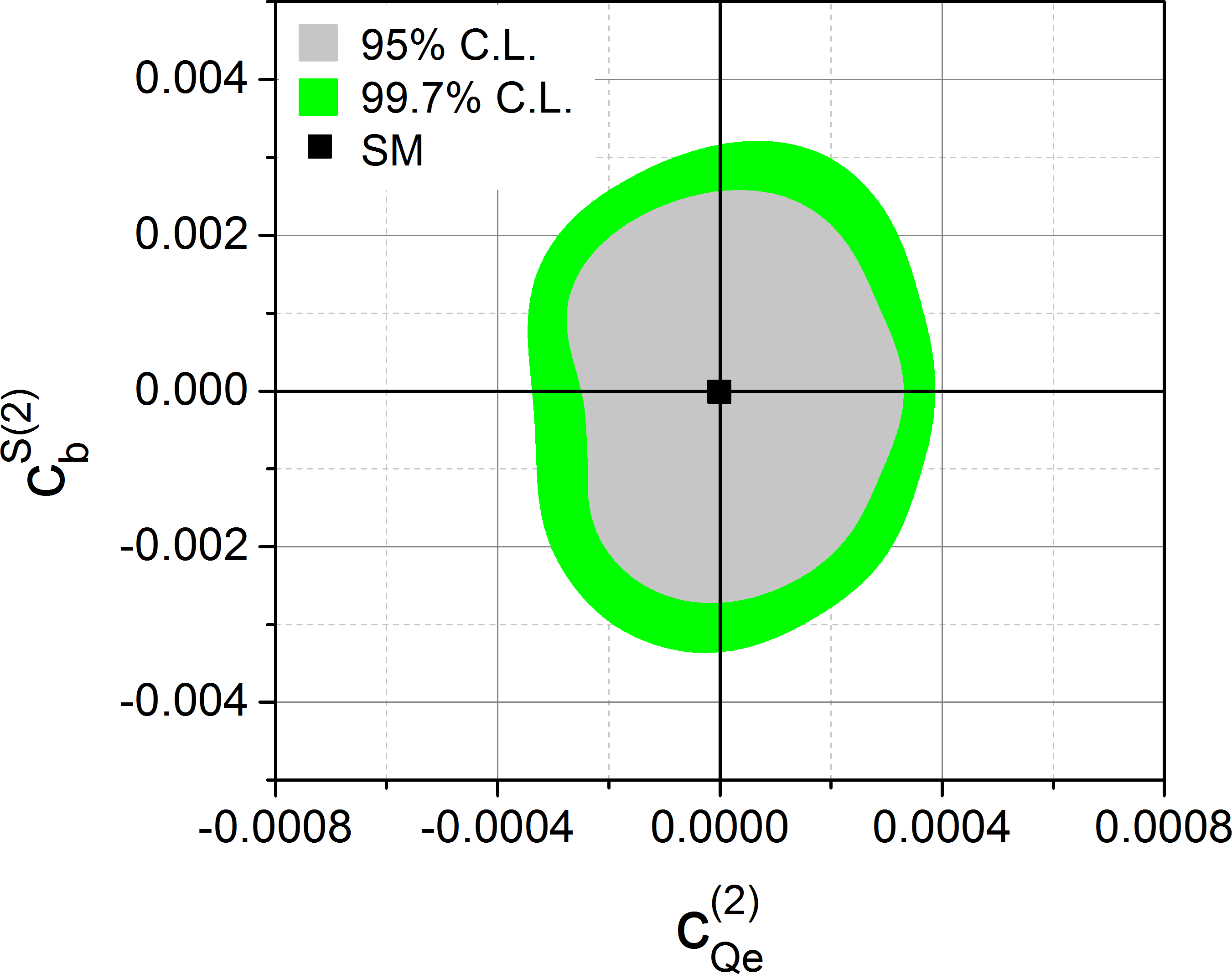}
\caption{}
\label{fig9:d}
\end{subfigure}\hspace{0.02\linewidth}
\begin{subfigure}{0.34\linewidth}
\includegraphics[width=\linewidth]{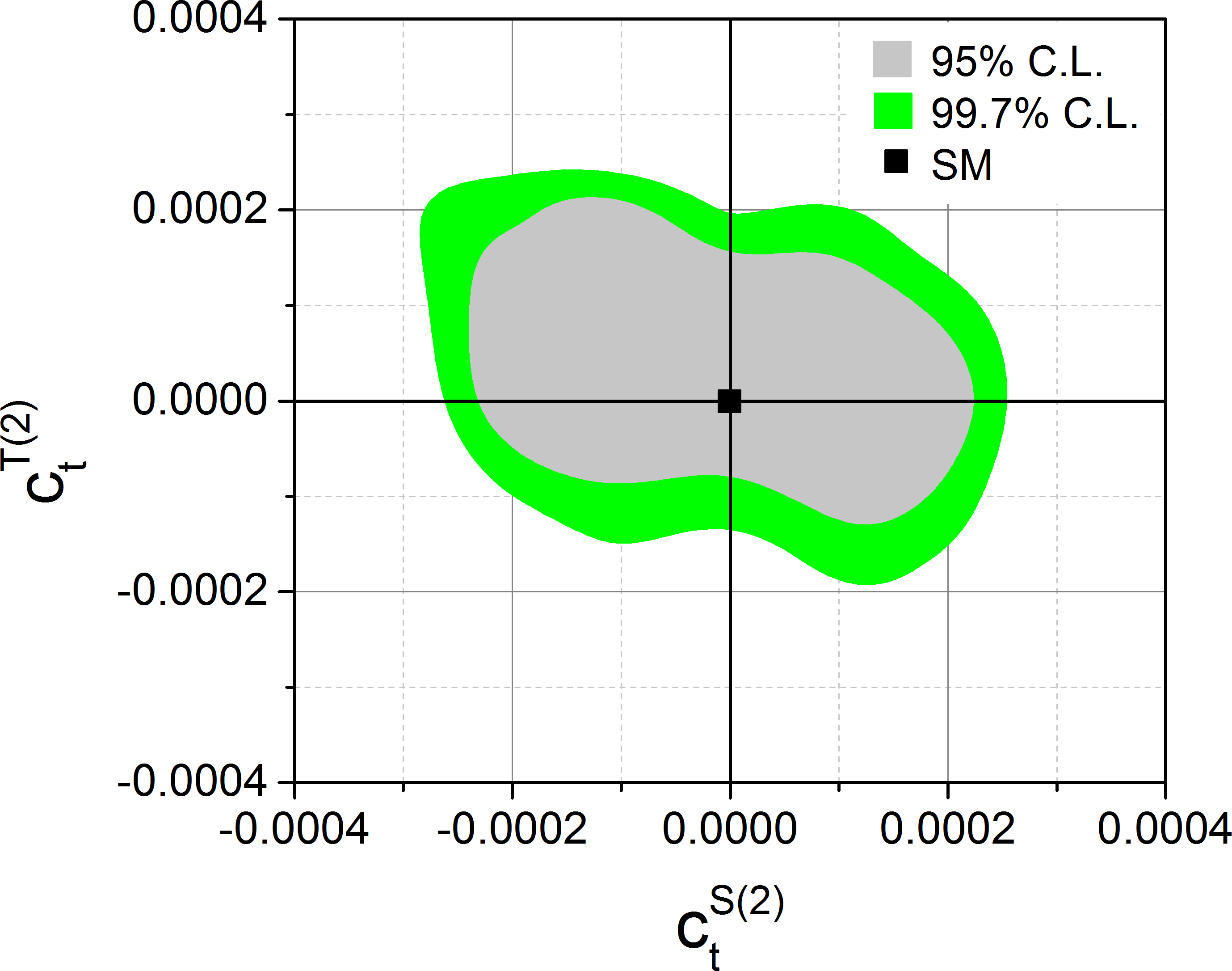}
\caption{}
\label{fig9:e}
\end{subfigure}\hfill
\caption{Two-dimensional 95\% and 99.7\% C.L. contours of (a) $c_{tW}$ - $c_{tZ}$, (b) $c_{tW}$ - $c_{bW}$, (c) $c_{Ql}^{-(2)}$ - $c_{Ql}^{3(2)}$, (d) $c_{Qe}^{(2)}$ - $c_b^{S(2)}$ and (e) $c_t^{S(2)}$ - $c_t^{T(2)}$ planes, with all other parameters fixed to zero.}
\label{fig9}
\end{figure}

Figure~\ref{fig9} presents the two-dimensional 95\% and 99.7\% C.L. contours in the $c_{tW}$ - $c_{tZ}$, $c_{tW}$ - $c_{bW}$, $c_{Ql}^{-(2)}$ - $c_{Ql}^{3(2)}$, $c_{Qe}^{(2)}$ - $c_b^{S(2)}$ and $c_t^{S(2)}$ - $c_t^{T(2)}$ planes, obtained from a two-parameter analysis. These two-dimensional likelihood scans are evaluated for the 90\% $b$-tagging efficiency at the 10 TeV muon collider. In this analysis, the 95\% and 99.7\% C.L. contours are derived by taking $\chi^2$ values of $5.99$ and $11.7$, respectively, corresponding to two degrees of freedom.

\section{Conclusions} \label{Sec5}

We have investigated the potential of a future muon collider for process $\mu^- \mu^+ \rightarrow Wtb$ at a center-of-mass energy of 10 TeV, accounting for realistic detector effects. We showed that the process $\mu^- \mu^+ \rightarrow Wtb$ is highly sensitive to the SMEFT operators $\leftindex^\ddagger {O}_{uW}^{(ij)}$, $\leftindex^\ddagger {O}_{dW}^{(ij)}$, $\leftindex^\ddagger {O}_{uB}^{(ij)}$, $O_{lq}^{1(ijkl)}$, $O_{lq}^{3(ijkl)}$, $O_{eq}^{(ijkl)}$, $\leftindex^\ddagger {O}_{lequ}^{1(ijkl)}$, $\leftindex^\ddagger {O}_{lequ}^{3(ijkl)}$ and $\leftindex^\ddagger {O}_{ledq}^{(ijkl)}$ on anomalous couplings of two heavy quarks and gauge boson(s) or two heavy quarks and two leptons. We present the kinematic distributions of the signal and relevant background processes using cut-based analysis to distinguish signal events from backgrounds. We determined 95\% C.L. limits for the Wilson coefficients of these operators using the $\chi^2$ test with the analysis of the $m_{\ell \ell bb}$ observable. The best limits in our study are obtained at the loose working point (90\% b-tagging efficiency); these are $[-0.01005; 0.00560]$ for the coefficient $c_{tW}$, $[-0.00347; 0.00486]$ for the coefficient $c_{tZ}$, $[-0.01848; 0.02271]$ for the coefficient $c_{bW}$, $[-0.00143; 0.00055]$ for the coefficient $c_{Ql}^{3(2)}$, $[-0.00018; 0.00019]$ for the coefficient $c_{Ql}^{-(2)}$, $[-0.00017; 0.00017]$ for the coefficient $c_{Qe}^{(2)}$, $[-0.00051; 0.00132]$ for the coefficient $c_b^{S(2)}$, $[-0.00004; 0.00004]$ for the coefficient $c_t^{T(2)}$, $[-0.00014; 0.00016]$ for the coefficient $c_t^{S(2)}$ at muon collider.

In Table~\ref{tab4}, we present a comparison between the 95\% C.L. limits on the Wilson coefficients obtained in this study and those reported in recent experimental studies \cite{Sirunyan:2021rfv,Hayrapetyan:2023yjz}. Among the experimental studies, the most precise limits for the Wilson coefficients were determined by the CMS collaboration \cite{Hayrapetyan:2023yjz} at a center-of-mass energy of $13$ TeV and an integrated luminosity of $138$ fb$^{-1}$. In our study, the limits obtained for the coefficients $c_{tW}$, $c_{tZ}$, $c_{bW}$, $c_{Ql}^{3(2)}$, $c_{Ql}^{-(2)}$, $c_{Qe}^{(2)}$, $c_t^{T(2)}$ and $c_t^{S(2)}$ at muon collider are approximately 56, 140, 36, 2662, 11162, 12235, 10000 and 18667 times more precise, respectively, than the limits reported in Ref.~\cite{Hayrapetyan:2023yjz}.

In light of previous SMEFT studies of top-quark interactions and the constraints established by the LHC experiments, our findings highlight the remarkable potential of the $\mu^- \mu^+ \rightarrow Wtb$ process as a complementary probe of physics beyond the SM. The projected sensitivities at a 10 TeV muon collider, particularly for two-heavy-quark-gauge boson(s) and two-heavy-quark-two-lepton operators, indicate substantial improvements over existing experimental bounds. These results demonstrate that the clean experimental environment and the high-energy reach of future muon colliders can provide unique opportunities to uncover deviations in third-generation quark interactions. Ultimately, the $\mu^- \mu^+ \rightarrow Wtb \rightarrow \ell^+ \nu_\ell b \ell^- \bar{\nu}_\ell \bar{b} $ channel offers a powerful pathway toward resolving the electroweak structure of new physics and extending our understanding of fundamental interactions beyond the SM.

\section*{Declaration of competing interest}

The authors declare that they have no known competing financial interests or personal relationships that could have appeared to influence the work reported in this paper.

\end{document}